\documentclass[aps,pra,reprint,superscriptaddress,longbibliography,amsmath,amssymb]{revtex4-2}

\usepackage{graphicx}
\usepackage{dcolumn}
\usepackage{bm}
\usepackage{xcolor}
\usepackage{amsmath,amsfonts,amssymb}
\usepackage{mathtools}
\usepackage{dsfont}
\usepackage{braket}
\usepackage[mathscr]{euscript}
\usepackage{amsthm}
\usepackage{comment}
\usepackage[ruled,vlined]{algorithm2e}
\usepackage{placeins}

\theoremstyle{plain}

\theoremstyle{definition}

\theoremstyle{remark}

\newcommand{\citationneeded}[1][]{\textsuperscript{\color{blue} [citation needed]}}

\DeclarePairedDelimiterX{\norm}[1]{\lVert}{\rVert}{#1}

\newcommand{\mathbbm}[1]{\text{\usefont{U}{bbm}{m}{n}#1}} 
\newcommand{\Herm}{{\rm Herm}}

\newcommand{\im}{\mathrm{i}\mkern1mu}
\DeclareMathOperator{\id}{{\mathbbm{1}}}

\DeclareMathOperator{\ran}{\text{Img}}
\DeclareMathOperator{\rank}{\text{Rank}}
\DeclareMathOperator{\linspan}{\text{Span}}

\DeclareMathOperator{\vecop}{vec}

\newcommand{\Ndim}{n}

\newcommand{\underrel}[2]{\mathrel{\mathop{#2}\limits_{#1}}}

\UseRawInputEncoding

\usepackage{hyperref}
\hypersetup{
    colorlinks,
    linkcolor={red!50!black},
    citecolor={blue!50!black},
    urlcolor={blue!80!black}
}

\begin{document}

\title{Identifiability of Autonomous and Controlled Open Quantum Systems}

\author{Waqas Parvaiz}
\affiliation{%
Department of Computer Science,
 Czech Technical University in Prague, Czech Republic
}

\author{Johannes Aspman}
\affiliation{%
Department of Computer Science,
 Czech Technical University in Prague, Czech Republic
}%

\author{Ales Wodecki}
\affiliation{%
Department of Computer Science,
 Czech Technical University in Prague, Czech Republic
}%

\author{Georgios Korpas}
\affiliation{%
Quantum Technologies Group, The Hongkong and Shanghai Banking Corporation Limited, Singapore
}%
\affiliation{%
Department of Computer Science,
 Czech Technical University in Prague, Czech Republic
}
\affiliation{%
Archimedes Research Unit on AI, Data Science and Algorithms, Marousi, Greece
}

\author{Jakub Marecek}%
\affiliation{%
Department of Computer Science,
 Czech Technical University in Prague, Czech Republic
}%

\begin{abstract}

Open quantum systems are a rich area of research {at} the intersection of quantum mechanics and stochastic analysis. {Starting from the Gorini--Kossakowski--Sudarshan--Lindblad master equation and passing to the coherence-vector representation of the measurement dynamics, we connect autonomous and controlled open quantum systems to classical linear and bilinear system identification theory.} This allows us to formulate corresponding notions of identifiability and parameter reconstructability for these systems, taking as input the state estimates produced by quantum state tomography, and providing conditions under which the probed quantum system is reconstructible. By the uniqueness part of the Gorini--Kossakowski--Sudarshan theorem, the master-equation parameters are always determined by the system matrices of these linear and bilinear systems; we give this map explicitly in terms of structure constants, together with its conditioning, so that identifiability of the open system reduces to that of the corresponding linear or bilinear system. We give conditions under which the system matrices, and hence the master equation, can be reconstructed from sampled data. Numerical experiments on a driven two-qubit system show that exact reconstruction sets in once these conditions are met, and that driving enables accurate identification from a simple initial state in the presence of shot noise. We set the groundwork for a number of constructive approaches to the identification of open quantum systems.
\end{abstract}

\maketitle

\section{Introduction}\label{sec:intro}
Quantum technologies are argued to have significant potential {across science and industry} \cite{Preskill2018}, and recent experiments demonstrate utility and early fault tolerance \cite{arute2019quantum,Kim2023,google2024_smaller}. These quantum devices are not isolated, and characterization of environmental noise processes is of considerable importance. {To fully understand a system, accurate estimates of its key features are required} \cite{Wiseman2009}. However, limited understanding of open quantum systems (OQS), both autonomous and controlled, hinders our ability to identify a model that captures their evolution.

{Identification} of a system, given suitable experimental design choices, usually leverages optimization algorithms which result in estimation of system parameters. Upon successful identification, we can then utilize {the identified model} to make predictions about future behavior of the system. For linear systems this theory is classical and consists of the identifiability of parametrized state-space models~\cite{bellman1970structural,glover1974parametrizations}, the persistent-excitation conditions on the data~\cite{astrom1971system,ljung1971characterization}, and the convergence of parametric estimators~\cite{ljung1978convergence} which were established in the 1970s. For bilinear systems, which arise once the quantum system is driven, realization theory dates from the same period~\cite{Isidori1973}, but conditions for identifiability from input--output data are much more recent~\cite{Sontag2009} and considerably less developed. Many of the past contributions on quantum system identification (parameter estimation) have focused on specific closed quantum systems that typically use traditional system identification methods such as the least squares method, or maximum-likelihood estimators \cite{ljung1999system, astrom1971system}. Some of the particular systems that have been studied include Hamiltonians with dipole moment operators controlled by laser pulses \cite{refId0}, quantum spin chains \cite{Sone2017}, and linear passive quantum systems coupled to bosons \cite{Gutta2016}. In a contribution on active learning of quantum Hamiltonians \cite{dutt2021active} the number of queries to the quantum processor needed to perform parameter identification up to a specified error is discussed. Hamiltonian learning that exploits structural assumptions, such as locality, can avoid full tomography altogether \cite{dasilva2011practical,huang2023learning}.
More general approaches have been proposed to identify closed quantum systems (under the assumption of identifiability) \cite{Wang_2020,bonnabel:hal-00447800, Burgarth_2012, Zhang2014}. Most notably, the authors of Ref. \cite{8022944} developed a two-step optimization algorithm for the Quantum Hamiltonian Identification (QHI) problem using quantum process tomography. In contrast, the literature on OQS identification has not been as well developed \cite{Zhang2015, xue2015,wallace2024learningdynamicsmarkovianopen,xue2021gradient, dumitrescu2020hamiltonian, popovych2024quantumopenidentificationglobal}. {In particular, Ref.~\cite{xue2021gradient} considers} a gradient algorithm, \cite{dumitrescu2020hamiltonian} consider least squares estimators using convex optimization, and \cite{popovych2024quantumopenidentificationglobal} consider subspace methods and semi-definite programming relaxations of polynomial optimization problems. There is also a recent high-level overview of learning and characterizing quantum states and dynamics in multi-qubit systems across different physical platforms \cite{gebhart2023learning}.

Before developing identification protocols, it is essential to examine identifiability. This concerns whether the underlying process can, in principle, be uniquely determined from the available data. Typically, this involves asking 1) whether the data we can obtain is sufficient to uniquely determine the underlying process, 2) how to certify correctness of an estimate, and 3) how to design experiments that guarantee such uniqueness. For closed quantum systems, several aspects of identifiability are discussed and summarized in \cite{Wang_2020}, where the authors give a precise quantitative meaning to the notion of quantum identifiability of closed quantum systems using the transfer function of linear dynamical systems. Further advances have been made by generalizing the similarity transformation approach \cite{Travis1981ONSI,VAJDA1989217,ljung1999system, astrom1971system}. The aforementioned approaches treat the output as available in continuous time. In most experiments, however, expectation values are estimated from projective measurements on many repetitions of the experiment, which yields samples at discrete times; continuous weak measurement \cite{Wiseman2009} is an alternative, at the cost of measurement back-action. Refs.~\cite{Zhang2014,Zhang2015} already estimate system matrices from such sampled time traces, with the eigensystem realization algorithm, and Ref.~\cite{popovych2024quantumopenidentificationglobal} fits open-system models to sampled data by global optimization. Here we ask instead when the continuous-time generator, and hence the master-equation parameters, are uniquely determined by sampled data, both for autonomous and for driven systems. The question of when an open system is identifiable at all has, by contrast, received comparatively little systematic attention. The closest result is \cite{Burgarth_2014}, which shows that the Lindblad generator is determined by input--output data only up to similarity transformations; a number of recent protocols reconstruct Lindbladians directly from measurement data \cite{Boulant2003,Bairey2020,Samach2022,Pastori2022,Onorati2023} without addressing identifiability as such. When the full state can be prepared and measured, recovering the generator reduces to process tomography at several times followed by a matrix logarithm~\cite{Boulant2003, mohseni2006direct, howard2006quantum}.

{The closest works to the present article are Refs.~\cite{Zhang2014,Zhang2015},
which also employ the coherence-vector representation and study when the parameters of a Markovian open system are determined by observable time
traces. Those works sample observable time traces, identify a realization with the eigensystem realization algorithm, and recover the Hamiltonian and Lindblad parameters by solving polynomial equations obtained by matching transfer functions, which may have several solutions; they treat uncontrolled systems and need only a few observables. Here we consider driven systems, give conditions on the sampling under which the matrix logarithm returns the generator, and show that once the measurement or preparation frame is fixed, $\vec\theta$ and $\gamma$ follow from the generator by an explicit linear map with known conditioning, so that no polynomial system has to be solved.

This is what our contributions presented in this article provide. Driving the system makes the measurement dynamics
\emph{bilinear} rather than linear; with full tomography and piecewise-constant pulses, its identification reduces to that of a few linear systems, one per pulse setting, and we give the corresponding rank and sampling conditions. Working with
discretely sampled data throughout, we obtain sufficient conditions in both
regimes:   for the autonomous case, a sampling interval below the aliasing limit or, without a bound on the frequencies, irregular sampling, and for the driven case,
uniform sampling together with the appropriate pulse family. We then construct
the linear map taking the master-equation parameters to the system matrix,
show that it is always invertible, which makes the uniqueness part of the Gorini--Kossakowski--Sudarshan theorem \cite{gorini1976completely} explicit, and analyze the conditioning of the
resulting reconstruction, which controls how tomographic noise propagates into
the recovered rates.   A numerical study of a driven two-qubit system confirms these conditions and shows when control is needed in practice.


} 

This article is structured as follows: Section \ref{sec2:background}
outlines the essential material on open quantum systems and system
identification. In Section \ref{sec:level4} we cast a controlled OQS
evolution as a bilinear dynamical system, and Section \ref{sec:level5} gives the identifiability and parameter-reconstruction
conditions for both the controlled and uncontrolled cases. Section~\ref{sec:examples} applies the framework to two coupled qubits with local dissipation. Simulated tomographic data show that exact reconstruction sets in once the rank condition is met, that a symmetric initial state such as $|00\rangle$ does not identify the system without control, and that under shot noise driving keeps the data informative, whereas autonomous data stop improving once the system has relaxed. Finally, Section~\ref{sec:algorithms} provides pseudocode for the procedures together with a conditioning-based stability analysis.

\section{Background}\label{sec2:background}
\subsection{Open Quantum Systems}

Although closed quantum system identification has been extensively discussed in the literature \cite{Wang_2020,fu:hal-01326060,bonnabel:hal-00447800,Burgarth_2012,8022944}, studies on open quantum system (OQS) identification have received significantly less attention. There, identification is complicated by the interaction with the environment (or bath), $B$. In this context, the evolution is often considered in terms of the density matrix $\rho_{S}$, which belongs to the reduced system $S$. The combined system and bath Hamiltonian, $H_{\rm{full}}${,} {is} represented as a tensor product sum
\begin{equation}\label{eq:hamil_full}
    \begin{aligned}
    H_{\rm{full}} &= H_{\rm S} \otimes I + I \otimes H_{\rm B} + H_{I}
    \end{aligned}
\end{equation}
with $H_{I}$ being the interaction Hamiltonian. This full system is closed, and hence $\rho_{\rm full}$   evolves according to the Liouville-von Neumann equation \cite{Breuer_Petruccione_2010a}. The reduced system, however, has non-unitary evolution and more consideration is needed. Many works \cite{Breuer_Petruccione_2010a, Lindar2020lecture, Rivas2011OpenQS} have given a detailed analysis of the various descriptions of the evolution of OQS. These descriptions range from general non-Markovian perturbative master equations such as the Nakajima--Zwanzig equation \cite{Nakajima1958, Zwanzig1960}, through microscopic models such as the Redfield equation \cite{REDFIELD19651}, to dynamical maps such as the Kraus map \cite{KrausBook} and non-completely positive maps \cite{Carteret2008BeyondCPMaps}.  However, the most well known and commonly used description is the Gorini--Kossakowski--Sudarshan--Lindblad (GKSL) master equation \cite{gorini1976completely,Lindblad1976}, which in an orthonormal operator basis reads

\begin{equation}\label{canonicalGKSL}
    \begin{aligned}
        \dot \rho_{\rm{S}}(t)=& -\im [H_{S}(t),\rho_{\rm{S}}(t)]\\&+
        \sum_{j,k=1}^{N^2-1}\gamma_{jk}(t)\left(F_j\rho_{\rm{S}}(t) F_k-\frac{1}{2}\{F_kF_j,\rho_{\rm{S}}(t)\}\right),
    \end{aligned}
\end{equation}
where the positive semidefinite Hermitian matrix $\gamma$, known as the Kossakowski (or decoherence) matrix \footnote{Its individual entries are sometimes referred to as ``decay rates''; strictly, the decay rates are its eigenvalues, obtained by diagonalizing $\gamma$, i.e.\ by a unitary change of the basis of jump operators, which gives the diagonal Lindblad form \cite{Lindblad1976}.}, collects the coefficients of the dissipative part of the generator in the basis $\{F_j\}$. {The operators $F_{j}$ form a complete orthonormal basis, satisfying}

\begin{equation}\label{basis_conditions}
\begin{aligned}
    & \{F_m\}_{m=1}^{N^2-1}, &F_0=\frac{1}{\sqrt{N}}\mathbbm{1}, \\
    & F_m=F_m^\dagger, &\text{Tr}(F_mF_n)=\delta_{mn}
\end{aligned}
\end{equation}



where $N$ is the dimension of the system considered{; for a register of $q$ qubits, $N=2^{q}$. Although this master equation is capable of accurately modeling many physical systems, {see, e.g., Refs.}~\cite{WallsMilburn2008,Leibfried2003,Gambetta2008}, it should be noted that many approximations are used to arrive at this equation \cite{stefanini2025lindblad}. Firstly, \textit{separability}{,} which assumes that the system is initially uncorrelated with the environment, which is justified in many physical scenarios and is valid if one has good control of the system (note that \cite{Colla2022} showed there is a class of initial correlation for which it is possible to derive a Lindblad equation). Secondly, the \textit{Born approximation} \cite{Breuer_Petruccione_2010a}, which assumes weak system--environment coupling, so that the dynamics can be truncated at second order in $H_{I}$ and the total state approximated at all times by $\rho_{\rm S}(t)\otimes\rho_{\rm B}$, with the environment state $\rho_{\rm B}$ essentially unaffected by the system. Another common assumption is known as the \textit{Markovian approximation} \cite{rivas2010markovian}}, in which we keep the system memoryless which is valid for timescales where the system dynamics are much longer than the relaxation times of the environment correlation functions. There is no consensus on an exact definition of quantum Markovianity and how it is quantified.{ We} encourage the reader to study \cite{LI20181} for a more detailed understanding. Finally, \textit{complete positivity and trace preservation} \cite{NielsenChuang, KrausBook} which ensure $\rho_{S}$ retains essential properties of a density matrix. This is required to represent a valid physical evolution. However, this has been a topic of debate \cite{Pechukas1994, Alicki1995, Pechukas1995, Shaji2005}. 

It should be noted that many physical systems are in fact non-Markovian \cite{Hall2014} and approximations will not allow them to be accurately studied. However,  numerous studies \cite{xue2015, xue2019modeling,chen2019markovian,campbell2018system, nurdin2023markovian, Pleasance2020, Megier2017, Budini2013} have instead taken another avenue of embedding the principal non-Markovian system of interest into an augmented system whose overall dynamics is Markovian. The augmented system is {easier to simulate, and from it the} dynamics of the principal system can be {recovered}, provided that the Hilbert space of the larger system has reasonable dimensions.

\subsection{Elements of system identification}\label{sec:sysid}

As discussed in the introduction, the two relevant systems are the autonomous linear dynamical system (LDS) and the general bilinear dynamical system (BDS), which we describe here more formally in the state space representation. Typically, {an} LDS can be written in the form

\begin{equation}\label{eq:LDS_CONT}
    \begin{aligned}
        &\dot{\vec{x}}(t) = \mathbf{A} \vec{x}(t) + \mathbf{B} \vec{u}(t) \\
        &\vec{y}(t) = \mathbf{C}\vec{x}(t) + \mathbf{D}\vec{u}(t),
    \end{aligned}
\end{equation}
where $\vec{x}(t) \in \mathbb{R}^{n}$ is the state vector, $\vec{y}(t) \in \mathbb{R}^{p}$ is the observed vector{, and} $\vec u(t) \in \mathbb{R}^m$ is the input {vector}. $\mathbf{A} \in \mathbb{R}^{n \times n}, \mathbf{B} \in \mathbb{R}^{n\times m}, \mathbf{C} \in \mathbb{R}^{p\times n}, \mathbf{D} \in \mathbb{R}^{p\times m}$ represent the system, {input, output,} and feed-through matrices{,} respectively. The feed-through matrix is often neglected for physical systems and hence we set $\mathbf{D} = 0$. Note that autonomous systems, which do not have any inputs, can still be written in this form by choosing the input to be the Dirac delta with the initial state as the nonzero value at $t=0$.  Next, a state-space realization is determined by its input--output behaviour up to a similarity transformation if it is minimal, i.e.\ controllable and observable \cite{Antsaklis97}; minimality is therefore the natural condition for identifying the system matrices themselves. Parameter identifiability, by contrast, is defined through the map from the parameters to the input--output behaviour (the transfer function, or equivalently the Markov parameters) and does not presuppose minimality; minimality enters only through the similarity-transformation test, via the result just mentioned \cite{glover1974parametrizations}. Known structure can thus remove the similarity freedom, and non-minimal systems can still be identifiable, as in the quantum settings of Refs.~\cite{Zhang2014,Zhang2015,Burgarth_2014}. A system is controllable if{, for any initial state $\vec{x}(0)$ and any target state $\vec{x}_{1}$, there exists an input $\vec{u}(t)$ steering the former to the latter in finite time. A} system is observable if one can uniquely determine the initial state from the output {$\vec y(t)$}. A practical way to determine the controllability and observability of a system, over a specified time interval, is to compute the rank of the corresponding controllability  and observability matrices, $\mathbf{CM} \in \mathbb{R}^{n\times nm}$ and $\mathbf{OM} \in \mathbb{R}^{np \times n}$,
respectively. These are defined as follows

\begin{equation}\label{Linear_OM_CM}
    \begin{aligned}
        \mathbf{OM}^{(\text{lin})} &= \begin{bmatrix}
            \mathbf{C} \\ \mathbf{CA} \\ \vdots \\ \mathbf{C}\mathbf{A}^{n-1}
        \end{bmatrix}\in\mathbb{R}^{np\times n}, \\[2pt]
        \mathbf{CM}^{(\text{lin})} &= \begin{bmatrix}
            \mathbf{B} & \mathbf{AB} & \cdots & \mathbf{A}^{n-1}\mathbf{B}
        \end{bmatrix}\in\mathbb{R}^{n\times nm}.
    \end{aligned}
\end{equation}
If $\mathbf{CM}^{(\text{lin})}$ ($\mathbf{OM}^{(\text{lin})}$) has full rank{,} then the system is controllable (observable) \cite{Sarachik_1997}. Note that, if the initial state may be arbitrary and $n=p$, the matrix $\mathbf B$ may be taken to be the identity. This leads to controllability being trivially satisfied. 

The second type of system naturally occurring in open quantum system identification is the bilinear dynamical system (BDS). More succinctly, the dynamics of the controlled quantum system can be written as
\begin{equation}\label{OQS_bilinear}
\begin{aligned}
    \dot{\vec{x}}(t) &= \mathbf{A}\vec{x}(t) + \sum_{j}\mathbf{N}_{j}u_{j}(t) \vec{x}(t),\\
    \vec{x}(0) &= \vec{b} = \mathbf{B}\vec{x}_{0},\\
    \vec{y}\left(t\right) &=\mathbf{C}\vec{x}\left(t\right).
\end{aligned}
\end{equation}
Here, the variables are the same as in \eqref{eq:LDS_CONT} with the addition of the bilinear coupling matrices, $\mathbf{N}_j \in \mathbb{R}^{n,n}$. We note that this is a BDS of type II in the language of \cite{Sontag2009}, where the initial state is fixed{, $\vec x(0) = \vec b$}.
Here $\mathbf{B}$ encodes the state-preparation map and{,} as in the linear case{,} {enters} only through the initial condition, which is why the direct driving term $\mathbf{B}\vec{u}$ of the general bilinear model is absent.
The presence of $\mathbf{N}_{j}$ requires us to modify these definitions of controllability and observability \cite{Isidori1973, Sontag2009}. In the case of the BDS \eqref{OQS_bilinear}, this is given by the criteria
\begin{equation}\label{eq:BDS_reach_and_obs}
\begin{aligned}
\linspan\{\mathbf{W}\vec{b} : \mathbf{W}\in\mathcal{W}\} &= \mathbb{R}^{n},\\
\linspan\{\vec{c}_q^{\,T}\mathbf{W} : \mathbf{W}\in\mathcal{W},\ q\le p\} &= \mathbb{R}^{1\times n},
\end{aligned}
\end{equation}
respectively. Here $\mathcal{W}$ is the set of all products $\mathbf{A}_{i_1}\cdots\mathbf{A}_{i_k}$ of $0\le k\le n-1$ factors $\mathbf{A}_{i_l}\in\{\mathbf{A},\mathbf{N}_1,\ldots,\mathbf{N}_m\}$, with the identity for $k=0$, $m$ is the number of controls, and $\vec{c}_q^{\,T}$ is the $q$-th row of the observation matrix $\mathbf{C}$. We denote by $\mathbf{CM}^{(\text{bi})}$ the matrix whose columns are the vectors in the first set and by $\mathbf{OM}^{(\text{bi})}$ the matrix whose rows are the vectors in the second set; the criteria are that both have full rank.


\section{A Dynamical-Systems View of Open Quantum Systems}\label{sec:level4}

\subsection{Controlled Open Quantum Systems as bilinear systems}\label{sec:bds}
As mentioned earlier, the dynamics of most open quantum systems can be reduced (or embedded) to the GKSL master equations, \eqref{canonicalGKSL}. In this section, we will show how to represent this evolution as a dynamical system. We restrict the analysis to a time-independent drift Hamiltonian and Kossakowski matrix; the only time dependence is through the control amplitudes $u_k(t)$. A time-dependent $H_0(t)$ or $\gamma(t)$ leads to the same equations with time-dependent $\mathbf{A}(t)$ and $\vec\beta(t)$, i.e.\ a linear time-varying system, whose identifiability is not covered by the results below.

In practice, the dynamics of quantum systems are obtained by observing their response after probing with an input control pulse \cite{DAlessandro2021}. The controllability and development of control strategies for open quantum systems is an ongoing field of study with limited understanding \cite{Koch2016}. Since controls {affect} the dynamics of a quantum system, they are introduced via an additional term in the Hamiltonian
\begin{equation}\label{control_hamiltonian}
   H(t) = H_{0} + \sum_{k=1}^{m} u_{k}(t)H_{k}.
\end{equation}
In a control theory context, the term $H_{0}$ is known as the drift Hamiltonian and represents the uncontrolled dynamics, and $H_{k}$ are the control Hamiltonians whose amplitudes are controlled via the (potentially time-varying) amplitude functions $u_{k}(t) \in \mathbb{R}$. 

We consider here Hermitian traceless Hamiltonians. This is a reasonable setup since the nonzero
trace part leaves the dynamics unchanged up to a global phase, and removing it
fixes the ambiguity in the GKSL generator. Such Hamiltonians lie in the real vector space of traceless Hermitian $N\times N$ matrices, of dimension $n\coloneqq N^{2}-1$. Taking $\Delta=\{F_{j}\}_{j=1}^{n}$ from \eqref{basis_conditions} as an orthonormal basis of that space, equivalently a basis of $\mathfrak{su}(N)$ up to an overall factor of $\im$, every such Hamiltonian admits a unique expansion; for the drift term in \eqref{control_hamiltonian},
\begin{equation}\label{eq:decomposition}
H_0=\sum_{j=1}^{n}\theta_jF_j,\qquad \theta_j=\mathrm{Tr}[F_jH_0]\in\mathbb{R}.
\end{equation}

We may collect the coefficients in
$\vec{\theta}=[\theta_{1},\dots,\theta_{n}]^{\top}\in\mathbb{R}^{n}$, which is
the vector to be identified.  Due to physical constraints of the system, only a small number $s$ of these coefficients is nonzero in practice, typically $s\ll n$ \cite{Zhang2014} but we retain the full vector
throughout, since the reconstruction of Sec.~\ref{sec:general_param_reconst}
inverts a map defined on all of $\mathbb{R}^{n}$. The anti-symmetric and symmetric structure constants, $f_{jkl}$ and $g_{jkl}$ respectively, are thus defined through  the commutation and anti-commutation relations
\begin{subequations}
    \begin{align}\label{eq:antisymmtric_structure_const}
        [F_{j}, F_{k}] =& \im\sum_{l=1}^{N^{2} - 1} f_{jkl}F_{l} \\ \label{eq:symmtric_structure_const}
        \{F_{j}, F_{k}\} =& \frac{2}{N}\delta_{jk}\id + \sum_{l=1}^{N^{2} - 1} g_{jkl}F_{l}.
    \end{align}
\end{subequations}
 The generators also satisfy the conditions \eqref{basis_conditions} of the jump operators, and so, without loss of generality, we can equate the operators in \eqref{canonicalGKSL} with this basis. Furthermore, the control Hamiltonians are known, so we expand them in the same basis with known coefficients $h_{kl}$,
\begin{equation}\label{reformulated_control_hamiltonian}
    H_k = \sum_{l=1}^{N^{2}-1} h_{kl}F_l, \qquad k = 1,\ldots,m .
\end{equation}

The reduced density operator $\rho_{\rm{S}}$ can also be decomposed in terms of these generators
\begin{equation}\label{eq:bloch_rho}
    \rho_{\rm{S}} = \frac{1}{N}\id_{N} +  \sum_{j=1}^{N^{2} - 1} x_{j}F_{j}.
\end{equation}
Here, $\rho_{\rm{S}}$ is parameterized by $N^{2} - 1$ elements $x_{j} := \mathrm{Tr}[F_{j}\rho_{\rm{S}}]$ of the coherence vector $\vec{x}$ \cite{Hioe1981}. Since each $F_{j}$ is an observable, these elements are simply their expectation values, which can be obtained by measurements. Physically, $x_j$ are coefficients which correspond to the average value obtained when measuring $F_{j}$ on an ensemble of identically prepared copies of $\rho_{\rm{S}}$. Hence, the state is encoded in the set of expectation values of a complete set of independent observables $\{F_{j}\}$.

Since the whole dynamics can be captured by $\vec{x}$, it represents the state and hence in the context of system identification it is called the state vector. By using the above, we can now show that a generic controlled open quantum system described by the GKSL equation \eqref{canonicalGKSL} can be recast into
a BDS form
as (we provide the full derivation in {Appendix}~\ref{ap:derivation_LDS})

\begin{equation}\label{eq:BLDS_lindbladian}
        \begin{aligned}
        \dot{\vec{x}}(t) = \underbrace{(\mathbf{A}^{(l)} + \mathbf{A}^{(d)})}_{\mathbf{A}}\vec{x}(t) + \vec{\beta} + \sum_{k=1}^{m}\mathbf{N}_{k}u_{k}(t)\vec{x}(t),
    \end{aligned}
    \end{equation}
where we have defined
\begin{equation}\label{eq:LDS_matrices_and_vectors}
    \begin{aligned}
        \mathbf{A}^{(l)}_{jk} &= -\sum^{N^{2} - 1}_{l=1} \theta_{l}f_{jkl}, \\
        \mathbf{A}^{(d)}_{jk} &= -\sum^{N^{2} - 1}_{l,m=1} \gamma_{lm}D^{(j,k)}_{lm}, \\
        \beta_{j} &= \frac{\im}{N}\sum^{N^2 - 1}_{k,l = 1} \gamma_{kl}f_{jkl},\\
        (\mathbf{N}_k)_{jl}&=-\sum_{r=1}^{N^{2}-1} h_{kr}f_{jlr},\\
        D^{(j,k)}_{lm} &= \frac{1}{4}\sum_{{r}=1}^{N^{2}-1} z_{l{r}k}f_{jm{r}} + \bar{z}_{m{r}k}f_{jl{r}}, \\
    z_{jkl} &= f_{jkl} + \im g_{jkl}.
    \end{aligned}
\end{equation}
{Here the summation index $r$ is a dummy, renamed so as not to clash with $n=N^{2}-1$.}
Here, $\mathbf{A}^{(l)}\in \mathbb{R}^{(N^{2} - 1) \times (N^{2} - 1)}$ is the system matrix coming from the Liouville-von Neumann dynamics of \eqref{canonicalGKSL}, while $\mathbf{A}^{(d)} \in \mathbb{R}^{(N^{2} - 1) \times (N^{2} - 1)}$, $\vec{\beta} \in \mathbb{R}^{N^2 - 1}$
are the system matrix and vector coming from the dissipative dynamics. Finally, each of the $m$ matrices $\mathbf{N}_{k}$ is in $\mathbb{R}^{(N^{2} - 1) \times (N^{2} - 1)}$.

Although $z_{jkl}$ and $D^{(j,k)}_{lm}$ are complex, the objects entering
\eqref{eq:BLDS_lindbladian} are necessarily real, as required by
$\vec{x}(t)\in\mathbb{R}^{{n}}$. Indeed
$\overline{D^{(j,k)}_{lm}}=D^{(j,k)}_{ml}$, and $\overline{\gamma_{lm}}=\gamma_{ml}$, so
$\overline{\mathbf{A}^{(d)}_{jk}}=-\sum_{l,m}\gamma_{ml}D^{(j,k)}_{ml}=\mathbf{A}^{(d)}_{jk}$, while anti-symmetry of
$f_{jkl}$ in $(k,l)$ gives
$\beta_{j}=-\tfrac{1}{N}\sum_{k,l}\operatorname{Im}(\gamma_{kl})f_{jkl}\in\mathbb{R}$.
In particular, $\vec{\beta}$ depends only on the imaginary part of $\gamma$ and
therefore vanishes identically whenever $\gamma$ is symmetric.

From the above, we see that the task of identifying the system amounts to finding the values of the parameters $\vec{\theta}$ of the Hamiltonian and the elements $\gamma_{jk}$ of the Kossakowski matrix. The matrix $\mathbf{A}^{(l)}$ is always antisymmetric, regardless of whether the system is open or closed. On the other hand, $\mathbf{A}^{(d)}$ in general does not have any symmetry properties. However, when $N = 2$ (e.g. a single qubit) or when the Kossakowski matrix is symmetric, $\mathbf{A}^{(d)}$ is symmetric as well. In the latter case, we also see that the vector $\vec\beta$ vanishes. The bilinear coupling matrices $\mathbf{N}_{k}$ do not need to be identified: by \eqref{eq:LDS_matrices_and_vectors} they are fixed by the antisymmetric structure constants and the coefficients $h_{kl}$ of the known control Hamiltonians. If the control Hamiltonians are only partly known, for instance because of miscalibrated drive amplitudes, their coefficients enter $\mathbf{N}_k$ exactly as $\vec\theta$ enters $\mathbf{A}^{(l)}$, and can be recovered in the same way, since $\mathbf{T}_1$ has full column rank. 

Note that \eqref{eq:BLDS_lindbladian} is not in the standard form of the bilinear dynamical equation \eqref{OQS_bilinear} above, due to the presence of the vector $\vec{\beta}$. However, it is straightforward to reformulate it into this desired form by appending $\vec{\beta}$ as an additional column to $\mathbf{A}$ and appending a scalar element to the state vector, $\vec{x}$. Finally, to keep the system matrix square, and for consistency, we also append a row to $\mathbf{A}$ and $\mathbf{N}_{j}$. Hence, one can arrive at the standard form by making the replacement
\begin{equation}\label{eq:standard_form}
    \begin{aligned}
        \mathbf{A} &\rightarrow \begin{bmatrix}
           \mathbf{A} & \vec{\beta} \\ \mathbf{0} & 0 \end{bmatrix},\\
           \mathbf{N}_{j} &\rightarrow \begin{bmatrix}
               \mathbf{N}_{j} & \mathbf{0} \\
               \mathbf{0} & 0 \end{bmatrix}, \\
           \vec{x} &\rightarrow \begin{bmatrix}
               \vec{x} & 1
           \end{bmatrix}^{\top}.
    \end{aligned}
\end{equation}
All matrices now have the dimension $N^{2} \times N^{2}$. To keep the notation light, we use the same symbols $\mathbf{A}$, $\mathbf{N}_k$ and $\vec x$ for the augmented quantities; which form is meant is clear from the dimension, $n$ or $n+1 = N^2$, and the reconstruction of Sec.~\ref{sec:parameter_reconstruction} always refers to the $n\times n$ block and to $\vec\beta$. Correspondingly, each prepared state contributes a column $(\vec x_0^{\top},1)^{\top}$ to $\mathbf{B}$, the constant coordinate is known and treated as measured, so that $\mathbf{C}$ stands for $\mathbf{C}\oplus1$, and the rank conditions \eqref{Linear_OM_CM} and \eqref{eq:BDS_reach_and_obs} are applied with $n+1$ in place of $n$.

\subsection{Identification of the Autonomous LDS Form}\label{sec:autonomous_case}

In statistics and system identification theory, much effort is focused \cite{west2006bayesian} on the identification of a system without any controls present. The identification of such an autonomous system is possible{, provided that the observability matrix} \eqref{Linear_OM_CM} {is of full rank and that the initial state $\vec{x}(0)$ is sufficiently rich, so that all modes (eigenvalues) of $\mathbf{A}$ appear in the output $\vec{y}(t)$}. The derivation of the system matrix follows the same steps as seen in Appendix \ref{ap:derivation_LDS} (without the need to consider Appendix \ref{ap:control_term}). Again, the embedding is done as described above $\eqref{eq:standard_form}$ to ensure that the system has the usual LDS form. \\

In practice, for large quantum systems one might only have access to a set of local observables $\{O_{j}\}$ and the output vector obtained would correspond to the expectation value of these observables  $\vec{y}(t) = [\langle O_{1}\rangle, \dots ]^{\top}$. Since observables are Hermitian, we can decompose these into the same basis of generators as before so that $O_{j} = \sum_{k=1}^{N^{2} - 1} o^{(j)}_{k}F_{k}$, hence the $j$'th row in the output matrix, $\mathbf{C}$, {consists of} the elements $o^{(j)}_{k}$. It is not required that all these basis generators appear in the decomposition of the set of observables, so we can collect the unique basis elements present in a set $\mathcal{M} = \{ F_{v_{1}}, \dots, F_{v_m} \}$, where $v$ is a vector of length $m$ such that each element identifies the basis matrix to include in $\mathcal{M}$.  For instance, if our observables are $O_{1} = o^{(1)}_1F_{1} + o^{(1)}_{3}F_{3}$ and $O_{2} = o^{(2)}_{3}F_{3} + o^{(2)}_{4}F_{4}$, then we have $m = 3$ and $\mathcal{M} = \{ F_{1}, F_{3}, F_{4} \}$.

The size of the system matrices can be too large for efficient computation, but only the dynamics of the measured observables is needed. In the Heisenberg picture, $\mathrm{d}\langle F_l\rangle/\mathrm{d}t = \mathrm{Tr}[\mathcal{L}^{\dagger}(F_l)\,\rho_{\rm S}]$, so $\langle F_l\rangle$ couples to $\langle F_k\rangle$ only if $A_{lk} = \mathrm{Tr}[F_k\,\mathcal{L}^{\dagger}(F_l)] \neq 0$. Which entries of $\mathbf{A}$, and of the $\mathbf{N}_j$ in the driven case, can be nonzero is fixed by the model class, i.e.\ by which components of $\vec\theta$ and $\gamma$ may be nonzero and by the control Hamiltonians $H_k$. Following Refs.~\cite{Wang_2020,Zhang2015}, which build on control theory for classical nonlinear systems \cite{Nonlinear_systems_shankar}, we iterate
\begin{equation}\label{eq:accessible_set}
\begin{aligned}
    G_0 &= \mathcal{M},\\
    G_{j} &= G_{j-1}\cup\{F_k : A_{lk}\not\equiv 0 \text{ for some } F_l\in G_{j-1}\},
\end{aligned}
\end{equation}
where $A_{lk}\not\equiv 0$ means that the entry of $\mathbf{A}$, or of some $\mathbf{N}_j$, is nonzero for generic parameter values in the model class. In practice it suffices to evaluate these matrices once at random parameter values on that support. This is the construction of Ref.~\cite{Zhang2015}, which applies the adjoint Lindbladian to the measured observables, evaluated on the generic sparsity pattern so that the result does not depend on the unknown parameter values; for closed systems it reduces to commutators with the Hamiltonian terms, as in Ref.~\cite{Wang_2020}. The iteration saturates after at most $n$ steps at the accessible set $G$, and by construction the components in $G$ obey a closed system of equations. The following pseudocode summarizes the procedure.

\begin{algorithm}[H]
\caption{Construction of the accessible set}
\KwIn{$\mathcal{M}$; sparsity pattern of $\mathbf{A}$ and of the $\mathbf{N}_j$}
\KwOut{Accessible set $G$}
$G \gets \mathcal{M}$, $G_{\rm old} \gets \emptyset$\;
\While{$G \neq G_{\rm old}$}{
    $G_{\rm old} \gets G$\;
    $G \gets G \cup \{F_k : A_{lk}\not\equiv 0 \text{ for some } F_l \in G\}$\;
}
\KwRet{$G$}
\end{algorithm}
Writing $\vec x_G$ for the components of $\vec x$ in $G$, and $\mathbf{A}_G$, $\vec\beta_G$ for the corresponding rows and columns of $\mathbf{A}$ and the $\mathbf{N}_k$ and entries of $\vec\beta$, the measured components obey
\begin{equation}\label{eq:reduced_lds}
   \dot{\vec x}_G = \mathbf{A}_G\,\vec x_G + \vec\beta_G + \sum_{k=1}^m u_k(t)\,(\mathbf{N}_k)_G\,\vec x_G .
\end{equation}
Only the parameters that enter $\mathbf{A}_G$ and $\vec\beta_G$ can be identified; parameters associated with inaccessible degrees of freedom remain undetermined and are irrelevant for predicting the measured outputs. For the model of Sec.~\ref{sec:examples}, for instance, measuring only $\sigma_z^1$ gives an accessible set of 6 of the 15 basis elements, in agreement with Ref.~\cite{Zhang2015}, whereas adding a $\sigma_x$ drive on qubit~1 makes it all 15. For the rest of the discussion we do not reduce the system in this way and consider the full dimensions.
\\

\subsection{Relation to the Liouville-space representation}

The GKSL equation \eqref{canonicalGKSL} is commonly written in Liouville space as
$\tfrac{d}{dt}\mathrm{vec}(\rho_S)=\hat{\mathcal{L}}\,\mathrm{vec}(\rho_S)$, with a complex matrix $\hat{\mathcal{L}}\in\mathbb{C}^{N^2\times N^2}$.
The coherence-vector form \eqref{eq:BLDS_lindbladian} is the same generator in a different basis. Since $\{F_0,F_1,\dots,F_n\}$ in \eqref{basis_conditions} is an orthonormal basis of $\mathbb{C}^{N\times N}$, the change of basis from $\{|i\rangle\langle j|\}$ is unitary, and since $\mathcal{L}$ preserves Hermiticity and trace, its matrix elements in this basis are real:
\begin{equation}
\big(\mathrm{Tr}[F_j\,\mathcal{L}(F_k)]\big)_{j,k=0}^{n}
=\begin{pmatrix}0&\vec 0^{\top}\\ \sqrt{N}\,\vec\beta&\mathbf{A}\end{pmatrix},
\end{equation}
which is \eqref{eq:standard_form} up to the ordering and scaling of the constant component $x_0=\mathrm{Tr}[F_0\rho_S]=1/\sqrt{N}$. The two descriptions are therefore equivalent, and $(\mathbf{A},\vec\beta)$ determines $\mathcal{L}$ and vice versa.

We work in the coherence-vector form because it puts the problem directly in the standard form of system identification:
(i) the state and system matrices are real and the trace constraint is eliminated, so the classical results for real linear and bilinear systems apply without modification;
(ii) the outputs are expectation values of the measured observables, linear in $\vec x$ with a known output matrix $\mathbf{C}$ (Sec.~\ref{sec:autonomous_case});
(iii) the controls enter through known bilinear matrices $\mathbf{N}_k$, so \eqref{eq:BLDS_lindbladian} is a bilinear system of the form \eqref{OQS_bilinear}, to which \eqref{eq:BDS_reach_and_obs} and the input classes of~\cite{Sontag2009} apply directly; and
(iv) the parameters $(\vec\theta,\gamma)$ enter $\mathbf{A}$ and $\vec\beta$ linearly through \eqref{eq:LDS_matrices_and_vectors}, so recovering them is a linear inverse problem (Sec.~\ref{sec:level5}).

\section{Identifiability of Open Quantum Systems}\label{sec:level5}

Now that we have cast the dynamics of open quantum systems as a bilinear dynamical system, we can discuss the aspects of identification of said systems, namely the identifiability and the reconstruction of system parameters. In particular, we are interested in identifying the system matrix $\mathbf{A}$ in \eqref{eq:BLDS_lindbladian} and not  $\mathbf{N}$ since it is determined solely by the structure constants, and therefore known. Additionally, we are interested in the reconstruction of the parameters of the GKSL equation \eqref{canonicalGKSL}. In this section, we develop identifiability conditions that may be used to identify quantum systems. We extend the work of \cite{Wang_2020, Zhou2012, Zhang2014, Zhang2015} by providing general conditions for the identifiability of quantum systems. We organize this section as follows. In Sec.~\ref{sec:parameter_reconstruction} we begin by assuming that the (embedded) system matrix $\mathbf{A}$ of the bilinear system \eqref{eq:standard_form} related to the GKSL equation is known as a result of some system identification procedure, for instance an algorithm using polynomial matrix inequalities, which we will aim to develop as an extension in an upcoming paper. Concretely, the data are sample averages of the measured observables over $S$ copies of $\rho_S(t_i)$, i.e.\ $\hat{\vec y}(t_i)=\mathbf{C}\vec x(t_i)+\vec e_i$ with $\vec e_i$ zero-mean and of covariance $O(S^{-1})$, where $\mathbf{C}=\id_{n}$ for full state tomography; by \eqref{eq:bloch_rho} these are equivalent to the populations and coherences of $\rho_S(t_i)$. The matrices $\mathbf{B}$ and $\mathbf{C}$ are fixed by the chosen preparations and observables, while $\mathbf{A}$ and $\vec\beta$ are estimated from $\{\hat{\vec y}(t_i)\}$ and then mapped to $(\vec\theta,\gamma)$. Two distinct questions therefore arise. The first is whether the data determine $(\mathbf{A},\vec\beta)$; this is a property of the experiment (preparations, observables, sampling, inputs) and is the classical persistent-excitation requirement~\cite{astrom1971system}, addressed in Sec.~\ref{sec:sum_identification}. The second is whether $(\mathbf{A},\vec\beta)$ determines $(\vec\theta,\gamma)$; this is a property of the parametrization alone, and we show in Sec.~\ref{sec:parameter_reconstruction} that it always does.

Note that from knowing the embedded system matrix, one can distinguish, without ambiguity, between $\mathbf{A}$ and $\vec{\beta}$ of \eqref{eq:BLDS_lindbladian} as they inhabit independent sub-matrices in \eqref{eq:standard_form}. With this, we show that the reconstruction of the corresponding GKSL equation is always possible, and unique once $\mathbf{A}$ is expressed in the coherence-vector frame (Sec.~\ref{sec:gauge}); a similarity-transformed system matrix yields equivalent measurement dynamics but, in general, not the physical parameters. Section~\ref{sec:sum_identification} summarizes the main results in compact form.
\\


\subsection{Reconstruction of Master Equation Parameters}\label{sec:parameter_reconstruction}
In this section, the general parameter reconstruction is derived (see Sec.~\ref{sec:general_param_reconst}). Following this, we show that the situation is somewhat simplified when the Kossakowski matrix is symmetric (Sec.~\ref{sec:param_reconst_sym}).\\

\subsubsection{General Parameter Reconstruction:}\label{sec:general_param_reconst}
Recalling the results of Sec.~\ref{sec:level4}, we note that the system matrix is of the form
\begin{equation}
\mathbf{A}(\vec\theta,\vec\gamma) = \mathbf{A}^{(l)}(\vec{\theta}) + \mathbf{A}^{(d)}(\vec{\gamma}).
\end{equation}
To simplify the notation, denote $N^2 - 1 = n$. In order to discuss the recovery of the parameters of the original GKSL equation, i.e., $\vec{\theta}$ and $\vec{\gamma}$, we consider the map that constructs the matrix of the system $\mathbf{A}$ using the parameters $\vec{\theta}$, $\vec{\gamma}$ and the structure constants $f_{jkl}$ and label it
\begin{equation}
\phi:\mathbb{R}^{n}\times\Herm(n)\rightarrow\mathbb{R}^{n\times n},
\end{equation}
with 
\begin{equation}\label{eq:parameter_reconstruction_map}
\phi\left(\vec{\theta},\vec{\gamma}\right)=\mathbf{A}^{(l)}(\vec{\theta})+\mathbf{A}^{(d)}(\vec{\gamma}).
\end{equation}
and where $\Herm(n)$ denotes the real vector space of Hermitian $n\times n$
matrices, $\dim_{\mathbb{R}}\Herm(n)=n^{2}$, so that the parameter counting
below is unchanged.

Additionally, we define a map $\hat{\phi}:\Herm(n)\rightarrow\mathbb{R}^{n}$, which produces the vector $\vec{\beta}$ when provided the decoherence parameters $\vec{\gamma}$ as
\begin{equation}
\hat{\phi}\left(\vec{\gamma}\right)=\left(\frac{\im}{N}\sum_{k,l=1}^{n}\gamma_{kl}f_{jkl}\right)_{j=1}^{n} = \vec{\beta}.
\end{equation}

In particular, we are interested in inverting $\phi$ and $\hat{\phi}$ jointly, which amounts to the unique reconstruction of the parameters $\vec{\theta}$ and $\vec{\gamma}$ from known values of $\mathbf{A}$ and $\vec{\beta}$. The following map is needed to formalize the vectorization process, which aids in formulating the reconstruction. Let $\Gamma$ be a bijective map
\begin{equation}\label{eq:bijectivemap}
    \Gamma :  \{1, 2, \cdots , n^{2} \} \rightarrow  \{ (j,k) : j,k \in \{1,\ldots,n\} \},
\end{equation}
this represents the indexing necessary to vectorize $\mathbf{A}^{(l)}, \mathbf{A}^{(d)}$ and $\gamma$. Next, define
\begin{equation}\label{eq:reconstrct_matrices}
\begin{aligned}
\mathbf{T}_{1}=& - \begin{pmatrix}f_{\Gamma(1)1} & \cdots & f_{\Gamma(1)n}\\
\vdots & \ddots & \vdots\\
f_{\Gamma(n^{2})1} & \cdots & f_{\Gamma(n^{2})n}
\end{pmatrix} \in \mathbb{R}^{n^2 \times n} \\
\mathbf{T}_{2}=& -\begin{pmatrix}D_{\Gamma(1)}^{(\Gamma(1))} & \cdots & D_{\Gamma(n^{2})}^{(\Gamma(1))} \\
\vdots & \ddots & \vdots\\
D_{\Gamma(1)}^{(\Gamma(n^{2}))} & \cdots & D_{\Gamma(n^{2})}^{(\Gamma(n^{2}))}
\end{pmatrix}\in \mathbb{C}^{n^2 \times n^2}.
\end{aligned}
\end{equation}

Furthermore, define
\begin{equation}\label{matrix_M}
\mathbf{M}=\left[\begin{array}{cc}
\mathbf{T}_{1} & \mathbf{T}_{2}\\
0 & -\frac{\im}{N}\mathbf{T}_1^{\top}
\end{array}\right]\in\mathbb{C}^{\left(n+n^{2}\right)\times\left(n+n^{2}\right)}.
\end{equation}
Although $\mathbf{M}$ has complex entries, it maps the physical parameters, $\vec\theta\in\mathbb{R}^n$ and Hermitian $\gamma$ (a real vector space of dimension $n^2$), to the real pair $(\vecop\mathbf{A},\vec\beta)\in\mathbb{R}^{n^2}\times\mathbb{R}^n$, and we regard it as a real-linear map between these two spaces of equal real dimension $n+n^2$. Its singular values are the same whether it is computed as the complex matrix \eqref{matrix_M} or in an orthonormal real basis of $\Herm(n)$.

The matrix $\mathbf{M}$ is invertible for every orthonormal basis of traceless Hermitian matrices (Appendix~\ref{ap:proof_thm_reconstruct_general}), so $\vec\theta$ and $\gamma$ are always uniquely determined by $\mathbf{A}$ and $\vec\beta$. This is the uniqueness part of the GKS theorem \cite{gorini1976completely}; maps from the generator to the Kossakowski matrix also appear in the canonical-form literature~\cite{Hall2014,tscherbul2024bringing}; what we add is the joint map $\mathbf{M}$ for $(\vec\theta,\gamma)$ in terms of structure constants, its inverse, and its conditioning. Since $\mathbf{M}$ is a bijection between spaces of equal real dimension (Appendix~\ref{ap:proof_thm_reconstruct_general}), every real pair $(\mathbf{A},\vec\beta)$ arises from some Hermitian $\gamma$: the GKSL form constrains the system matrices only through $\gamma\succeq0$. Identifying $(\vec\theta,\gamma)$ is therefore equivalent to identifying an affine linear or bilinear system without further structure, and the quantum-specific information lies in positivity and in known structure such as sparsity, which enter through the constrained reconstruction of Sec.~\ref{sec:param_reconst_sym} and the accessible set \eqref{eq:accessible_set}. The two blocks of $\mathbf{M}$ play different roles. $\mathbf{T}_1$ always has full column rank, since $\mathfrak{su}(N)$ has trivial center, so that $[H,\cdot\,]$ vanishes on traceless operators only for $H=0$. $\mathbf{T}_2$, in contrast, is in general rank deficient: its kernel consists of non-unital dissipators that leave $\mathbf{A}$ unchanged, and these are fixed by $\vec\beta$ through the lower block of $\mathbf{M}$. The singular values of $\mathbf{M}$, and hence its conditioning, do not depend on the choice of orthonormal basis.
{We stress that} the reconstruction map is linear and therefore does not enforce
the physicality constraint $\gamma\succeq0$ required for CP-divisible
evolution. This is in fact useful, since the reconstructed $\hat\gamma$ provides an independent validity certificate, in that $\hat\gamma\not\succeq0$ signals either a non-Markovian (or non-CP)
underlying evolution or an identification error exceeding the noise floor.
The two cases can be separated. If the true $\gamma\succeq 0$, Weyl's inequality gives $\lambda_{\min}(\hat\gamma)\ge-\|\hat\gamma-\gamma\|_2\ge-\|\vec \xi-\hat{\vec \xi}\|$, which is bounded by \eqref{eq:error_bound}; a negative eigenvalue exceeding this bound in magnitude is therefore incompatible with a Markovian (CP) generator, whereas smaller ones are inconclusive. Independently, a Markovian semigroup yields the same generator for every sampling interval $\Delta t_i$, so a systematic dependence of $\hat\gamma$ on $\Delta t_i$ beyond this bound indicates non-Markovian dynamics \cite{wolf2008assessing}, while estimation error alone produces only fluctuations within it.
When the negative eigenvalues are within this bound, one may project onto the positive semidefinite cone,
$\gamma^{\star}=\arg\min_{\gamma\succeq0}\lVert\gamma-\hat\gamma\rVert_{F}
=\sum_{i}\max(\lambda_{i},0)v_{i}v_{i}^{\dagger}$. This projection is
non-expansive and hence does not amplify the error bound of
\eqref{eq:error_bound}.

\subsubsection{Parameter Reconstruction for  Symmetric Kossakowski Matrix:}\label{sec:param_reconst_sym}

In some situations, it might happen that the Kossakowski matrix $\gamma$ is symmetric, that is, $\gamma_{jk}=\gamma_{kj}$ (note that this further implies that $\gamma$ is real, since it is also Hermitian). Such systems have been briefly discussed in \cite{gorini1978properties}{, and constitute a subclass of unital dynamics: $\vec{\beta}$ depends only on $\mathrm{Im}\,\gamma$, so a real $\gamma$ forces $\vec{\beta}=0$, although the converse need not hold, since $\gamma\mapsto\vec{\beta}$ has a nontrivial kernel. The two-qubit model of Sec.~\ref{sec:examples} lies in this class when $g^+_k = g^-_k$.}

If we assume the Kossakowski matrix to be symmetric, the coherence vector representation, \eqref{eq:BLDS_lindbladian}, simplifies substantially. From the definitions, we easily see that this assumption implies firstly that $\vec{\beta}=0$, and secondly that
\begin{equation}\label{symmeytric_a_d}
   \mathbf{A}^{(d)}_{jk}=-\frac{1}{2}\sum_{l,m,p=1}^{n}\gamma_{lm}f_{jmp}f_{klp}\eqqcolon-\sum_{l,m}\gamma_{lm}\tilde{D}_{lm}^{(j,k)}.
\end{equation}
Because of the sum over $l,m$ and the symmetry of the Kossakowski matrix, $\mathbf{A}^{(d)}$ will necessarily be symmetric as well. Note also that, since $\vec{\beta}=0$, we do not need to consider the augmented system matrix as in \eqref{eq:standard_form}.
Since $\mathbf{A}^{(l)}$ is always anti-symmetric, from its definition, this means that we can easily distinguish the two components of the identified system matrix $\mathbf{A}$ as simply corresponding to its symmetric and anti-symmetric parts. Furthermore, the symmetry of the Kossakowski matrix means that there are at most $n(n+1)/2$ unique unknown elements in the vector $\vec{\gamma}$, rather than $n^{2}$. This allows for a transformation that reduces the column size of $\mathbf{T}_2$. Let $\vec\gamma_{\rm sym}\in\mathbb{R}^{n(n+1)/2}$ collect the independent entries of a symmetric $\gamma$, ordered by the bijection
\begin{equation*}
    \Gamma_{\rm sym} : \{1,\ldots,n(n+1)/2\} \to \{(l,m) : 1\le l\le m\le n\},
\end{equation*}
so that $(\vec\gamma_{\rm sym})_i = \gamma_{\Gamma_{\rm sym}(i)}$. Let $\mathbf{P}_{\rm sym}\in\mathbb{R}^{n^2\times n(n+1)/2}$ be the matrix with $\vecop\gamma = \mathbf{P}_{\rm sym}\vec\gamma_{\rm sym}$; it copies each $\gamma_{lm}$ with $l<m$ to the positions $(l,m)$ and $(m,l)$, and each diagonal entry to $(l,l)$. Then
\begin{equation}\label{eq:T_3}
    \vecop\mathbf{A}^{(d)} = \mathbf{T}_3\,\vec\gamma_{\rm sym},
    \qquad
    \mathbf{T}_3 = \mathbf{T}_2\,\mathbf{P}_{\rm sym},
\end{equation}
with $\mathbf{T}_3\in\mathbb{R}^{n^{2}\times n(n+1)/2}$, which we call the \emph{symmetric reduction} of $\mathbf{T}_2$. Each column of $\mathbf{T}_3$ is either a column of $\mathbf{T}_2$ (diagonal entries) or the sum of the two columns of $\mathbf{T}_2$ belonging to $(l,m)$ and $(m,l)$; equivalently, $\mathbf{T}_2$ may be replaced by the real tensor $\tilde D$ of \eqref{symmeytric_a_d}.

The symmetric reduction of $ {\mathbf{T}_{2}}$ will become important when checking whether one can recover the original GKSL equation from a linear system, which completes the identification process. We join together the aforementioned observations along with the uniqueness of the symmetric-antisymmetric decomposition (Toeplitz decomposition) to develop the following key result.\\

 {Assuming that} the Kossakowski matrix $\gamma$ is symmetric, the parameters $\vec{\theta}$ and $\vec{\gamma}$ are uniquely reconstructible from the values of $\mathbf{A}$ since $\mathbf{T}_{1}$ and  $\mathbf{T}_{3}$ have full column rank (see below) with,
\begin{equation}\label{eq:image_condition}
\begin{pmatrix}A_{\Gamma(1)}^{\left(l\right)}\\
\vdots\\
A_{\Gamma(n^{2})}^{\left(l\right)}
\end{pmatrix}\in\ran\mathbf{T}_{1},
\begin{pmatrix}A_{\Gamma(1)}^{\left(d\right)}\\
\vdots\\
A_{\Gamma(n^{2})}^{\left(d\right)}
\end{pmatrix}\in\ran\mathbf{T}_{3},
\end{equation}
where $A^{(l)}$ and $A^{(d)}$ may be computed using the symmetric decomposition
\begin{eqnarray}\label{eq:toeplitz_Decomposition}
\mathbf{A}^{(d)}=\frac12(\mathbf A+\mathbf{A}^{\top}),\quad \mathbf{A}^{(l)}=\frac12(\mathbf A-\mathbf{A}^\top).
\end{eqnarray}

Both $\mathbf{T}_1$ and $\mathbf{T}_3$ always have full column rank: $\mathbf{T}_1$ as in Sec.~\ref{sec:general_param_reconst}, and $\mathbf{T}_3$ (rank $n(n+1)/2$) because for symmetric $\gamma$ we have $\vec\beta=0$, so invertibility of $\mathbf{M}$ implies that $\mathbf{A}$ alone determines $(\vec\theta,\gamma)$. For exact data the range conditions \eqref{eq:image_condition} then hold automatically. Note that
$\mathbf{T}_{3}\in\mathbb{R}^{n^{2}\times n(n+1)/2}$ is rectangular for every
$n\geq 2$, so it is never invertible in the literal sense, and the relevant
object is the left inverse
$\mathbf{T}_{3}^{+}=(\mathbf{T}_{3}^{\top}\mathbf{T}_{3})^{-1}\mathbf{T}_{3}^{\top}$.\\\\

 { One can then discuss} the possibility of transferring known physical constraints on the parameters of the system $\theta$ and the decoherence rates $\gamma$. Using the result we proved for $\mathfrak{su}(2^{ {q}})$ in Appendix \ref{ap:struct_const_orthog}, one can see that the expression determining the off-diagonal elements reads
\begin{equation}
\mathbf{A}_{jk}^{(l)}=\sum_{p=1}^{n}\theta_{p}f_{pkj}=\theta_{p_{0}}f_{p_{0}\left(kj\right)kj},
\end{equation}
where $p_{0}\left(kj\right)$ is the unique non-zero index for which $f_{p_{0}kj}\neq0$. Therefore, imposing a constraint on the quantum system parameters $\theta_{p}$ can be done easily using the following
\begin{equation}
c_{1}\leq\theta_{p_{0}}\leq c_{2}\Leftrightarrow c_{1}f_{p_{0}\left(kj\right)kj}\leq \mathbf{A}_{jk}^{(l)}\leq c_{2}f_{p_{0}\left(kj\right)kj},
\end{equation}
when the structure constant
$f_{p_{0}\left(kj\right)kj}$
is positive. One need only reverse the inequalities when
$f_{p_{0}\left(kj\right)kj}$
is negative.
For the Kossakowski matrix, the entries of $\mathbf{A}^{(d)}$ are in general sums over several entries of $\gamma$, so bounds on individual entries of $\gamma$ do not translate into bounds on individual entries of $\mathbf{A}$. Such constraints can nevertheless be imposed directly in the reconstruction: positivity, $\gamma\succeq0$, and affine bounds on the entries of $\vec\theta$ and $\gamma$ are convex, so the constrained reconstruction
\begin{equation}
    \min_{\vec\theta,\;\gamma\succeq 0}\;
    \big\|\mathbf{M}(\vec\theta,\vecop\gamma)-(\vecop\hat{\mathbf{A}},\hat{\vec\beta})\big\|
\end{equation}
is a semidefinite program, to which further affine bounds can be added.

\subsubsection{The role of the measurement frame}\label{sec:frame}
\label{sec:gauge}

Classical identification procedures return a realization of the
 {input--output} map, not a distinguished matrix triple: any
$(\hat{\mathbf A},\hat{\mathbf B},\hat{\mathbf C})
 =(\mathbf{T}\mathbf{A}\mathbf{T}^{-1},\,\mathbf{T}\mathbf{B},\,\mathbf{C}\mathbf{T}^{-1})$
with $\mathbf T$ invertible
produces identical data. This is the sense in which the Lindblad
generator is determined by  {input--output} data only up to a similarity
transformation, as shown in
\cite{Burgarth_2014}.
The reconstruction maps $\phi$ and $\hat{\phi}$ introduced above,
however, are defined with respect to the canonical coherence-vector
basis $\{F_j\}_{j=1}^{n}$: applying $ {\mathbf{M}}^{-1}$ to a similarity-transformed
system matrix would yield parameters $(\vec{\theta},\gamma)$ that differ
from the physical ones and need not correspond to a valid GKSL
generator at all. Master-equation reconstruction is therefore well
posed only once the frame is fixed. \\

In our setting, the frame is fixed by the experimental design itself,
and the residual similarity freedom of \cite{Burgarth_2014} is
removed for two reasons. First, the output matrix $ {\mathbf{C}}$ is not estimated but
chosen: its rows are the known expansion coefficients $o^{(j)}_k$ of
the measured observables in the basis $\{F_j\}$, so $ {\mathbf{C}}$ is a known,
full-rank matrix. Second, the preparation map $\mathbf{B}$ is known. For a minimal realization, any triple producing the same input--output map is related to $(\mathbf{A},\mathbf{B},\mathbf{C})$ by a unique similarity transformation $\mathbf{T}$ \cite{Sontag1998}, which must satisfy $\mathbf{T}\mathbf{B}=\mathbf{B}$ and $\mathbf{C}\mathbf{T}^{-1}=\mathbf{C}$. Either condition alone forces $\mathbf{T}=\id$: the second if $\mathbf{C}$ has full column rank $n$ (informationally complete tomography), the first if $\mathbf{B}$ has full row rank $n$ (a spanning set of preparations). In Sec.~\ref{sec:examples} a single state is prepared, $\mathbf{B}=\vec b$, and it is the known $\mathbf{C}$ that fixes $\mathbf{T}$. Consequently the identified system matrix coincides with
$\mathbf{A}$ in the canonical coherence-vector frame, and the inversion of
$\mathbf{M}$ (respectively $\mathbf{T}_1$, $\mathbf{T}_3$) in this section is
unambiguous.

Note that this argument presupposes that $\mathbf{B}$ or $\mathbf{C}$ is known exactly and has full rank, i.e.\ that state preparation and measurement (SPAM) are free of errors. With SPAM errors the similarity freedom reappears as the gauge freedom of gate-set tomography \cite{nielsen2021gate}, and the reconstructed parameters are only as accurate as the calibration of preparation and measurement; we quantify this in Eq.~\eqref{eq:spam_bias}. When only some observables are measured, $\mathbf{C}$ does not have full column rank. If $\mathbf{B}$ does not have full row rank either, every invertible $\mathbf{T}$ with last row $(0,\ldots,0,1)$, $\mathbf{T}\mathbf{B}=\mathbf{B}$, $\mathbf{C}\mathbf{T}=\mathbf{C}$ and $\mathbf{T}\mathbf{N}_k=\mathbf{N}_k\mathbf{T}$ for all $k$ turns $\mathbf{A}$ into $\mathbf{T}\mathbf{A}\mathbf{T}^{-1}$ without changing any output. For a minimal system these are the only ambiguities \cite{ho1966effective,sussmann1976existence}, so $(\vec\theta,\gamma)$ is identifiable if and only if every such $\mathbf{T}$ commutes with $\mathbf{A}$. In the autonomous case this always fails: for $\vec u\in\ker\mathbf{C}$ with zero last entry and $\vec v$ orthogonal to the columns of $\mathbf{B}$, $\mathbf{T}=\id+\epsilon\,\vec u\vec v^{\,T}$ is such a transformation, and if all of these commuted with $\mathbf{A}$, then $\ker\mathbf{C}$ would contain an eigenvector of $\mathbf{A}$, contradicting observability. Identification from partial observations therefore requires additional information, such as the known structure of $H$ and of the jump operators, which restricts the admissible $\mathbf{A}$ as quantified by the accessible set \eqref{eq:accessible_set}, or control, which adds the constraints $\mathbf{T}\mathbf{N}_k=\mathbf{N}_k\mathbf{T}$. A general treatment is left for future work.

\subsection{Main Identifiability Results}\label{sec:sum_identification}

In this section, we summarize the preceding developments in two results, for autonomous and for driven systems. An experiment fixes the preparations (the columns of $\mathbf{B}$), the measured observables (the rows of $\mathbf{C}$), the admissible inputs $\vec u(t)$ and the sampling times $t_k$. We call the model \emph{identifiable} at $(\vec\theta,\gamma)$ if every $(\vec\theta',\gamma')$ that produces the same samples $\vec y(t_k)$ for all preparations and admissible inputs satisfies $(\vec\theta',\gamma')=(\vec\theta,\gamma)$. Identifiability of $(\mathbf{A},\vec\beta)$ is defined in the same way and, since $\mathbf{M}$ is invertible, is equivalent. The results concern noise-free data; finite samples are treated in Sec.~\ref{sec:examples}. The rank conditions below are polynomial in the entries of $\mathbf{A}$, which are linear in $(\vec\theta,\gamma)$. On any linear subspace of parameters, such as a fixed sparsity pattern, they therefore fail either everywhere or only on a set of measure zero, so verifying them at a single parameter value establishes generic identifiability \cite{bellman1970structural}.

\subsubsection{Autonomous Open Quantum System Identification}\label{sec:autonomous_id}
As we saw in the sections leading up to this one, the identification of autonomous open quantum systems reduces to  {the identification of an} LDS.

 {Before proceeding, an important remark on the sampling of the system is in order.} In \cite{Wang_2020,Burgarth_2012, Burgarth_2014,wallace2024learningdynamicsmarkovianopen,popovych2024quantumopenidentificationglobal} the continuous time identification of quantum systems is discussed, however, practically we only  have access to discrete time data. The recovery of continuous time systems from discrete time data has been studied in \cite{chen20,Ding2009}. In particular, a single discrete system sampled at a uniform rate does not in general determine the continuous system uniquely, since the matrix logarithm is not unique \cite{gonzalez2023, cubitt2012complexity}. Uniqueness is restored if the sampling interval obeys $|\mathrm{Im}\,\lambda(\mathbf{A})| < \pi/\Delta t$, in which case the principal logarithm returns the generator. This condition can be checked without knowing $\mathbf{A}$: for $\gamma\succeq0$, every eigenvalue of the generator satisfies
\begin{equation}\label{eq:alias_bound}
|\mathrm{Im}\,\lambda|\le 2\big(\|H\|+\mathrm{Tr}\,\gamma\big),
\end{equation}
with $\|\cdot\|$ the operator norm and $H$ including any constant drive. Indeed, $\lambda$ lies in the numerical range of the generator with respect to the Hilbert--Schmidt inner product, to which the Hamiltonian part contributes a purely imaginary term of modulus at most $2\|H\|$, while the dissipative part, written with Lindblad operators $L_k$ obtained by diagonalizing $\gamma$, has norm at most $2\sum_k\|L_k\|^2$, which is at most $2\,\mathrm{Tr}\,\gamma$. A priori bounds on $\|H\|$ and on the total rate $\mathrm{Tr}\,\gamma$ thus give a sampling interval for which the condition holds for all candidate parameters. Without such a bound, one requires either non-uniform sampling or several experiments with different sampling rates \cite{Ding2009}. More specifically, let the sampling pattern repeat periodically, with each period split into sub-intervals of lengths $\Delta t_i = t_i - t_{i-1}$; at the level of eigenvalues, the data at interval $\Delta t_i$ determine each eigenvalue $\lambda$ of $\mathbf{A}$ only up to multiples of $2\pi\im/\Delta t_i$. With one more sub-interval than $\mathbf{A}$ has eigenvalues, a spurious eigenvalue consistent with all of them must be an alias of the same $\lambda$ for two intervals, $\lambda+2\pi\im k_i/\Delta t_i=\lambda+2\pi\im k_j/\Delta t_j$ with $k_i,k_j\neq0$, which requires a rational ratio $\Delta t_i/\Delta t_j=k_i/k_j$. Irrational ratios therefore determine the eigenvalues uniquely \cite[Lemma~4]{Ding2009}, and $\mathbf{A}$ then follows from the eigenvectors, or Jordan structure, of a one-step map whose interval is non-pathological, i.e.\ no two eigenvalues of $\mathbf{A}$ differ by a nonzero multiple of $2\pi\im/\Delta t_i$ \cite[Sec.~5]{Ding2009}. These arguments use only the maps $e^{\mathbf{A}\Delta t_i}$, which full tomography provides directly, so they apply without inputs as well as with inputs held constant between samples, the zero-order-hold setting of \cite{Ding2009}. With the bound \eqref{eq:alias_bound}, only finitely many $k_i$ are possible, so in finite precision it suffices that the ratios stay away from the finitely many rationals $k_i/k_j$ that these allow, by a margin set by the noise. This allows for the identification of the discrete LDS matrices, from which one can recover the system and control matrices of the continuous LDS.\\

\noindent\textbf{Result 1 (autonomous systems).} Consider the augmented system \eqref{eq:standard_form} without control, with known $\mathbf{B}$ and $\mathbf{C}$, sampled at $t_k=k\Delta t$. Suppose that
(i) either $\mathbf{C}$ has full column rank and $\mathbf{CM}^{(\text{lin})}$ has full rank, or $\mathbf{B}$ has full row rank and $\mathbf{OM}^{(\text{lin})}$ has full rank; and
(ii) every eigenvalue $\lambda$ of $\mathbf{A}$ satisfies $|\mathrm{Im}\,\lambda|<\pi/\Delta t$, for the true and all candidate parameters.
Then the samples $\vec y(t_k)$, $k=0,\ldots,n+1$, for all preparations determine $\mathbf{A}$, and hence $(\vec\theta,\gamma)$, uniquely.

\noindent\emph{Proof.} Let $\boldsymbol{\Phi}=e^{\mathbf{A}\Delta t}$, so that $\vec x(t_k)=\boldsymbol{\Phi}^k\vec x(0)$. By (ii), $\mathbf{A}\Delta t$ is the principal logarithm of $\boldsymbol{\Phi}$, the unique logarithm whose eigenvalues lie in the strip $|\mathrm{Im}\,z|<\pi$ \cite{higham2008functions}. As a primary matrix function it is a polynomial in $\boldsymbol{\Phi}$, so $\mathbf{A}$ and $\boldsymbol{\Phi}$ generate the same Krylov subspaces, and the rank conditions in (i) hold with $\boldsymbol{\Phi}$ in place of $\mathbf{A}$. In the first case of (i), the states $\vec x(t_k)$ follow from $\vec y(t_k)$ via $\mathbf{C}^{+}$. Let $\mathbf{X}_-$ and $\mathbf{X}_+$ collect $\vec x(t_k)$ over all preparations for $k=0,\ldots,n$ and $k=1,\ldots,n+1$, respectively. Then $\mathbf{X}_+=\boldsymbol{\Phi}\mathbf{X}_-$ with $\mathbf{X}_-$ of full row rank, so $\boldsymbol{\Phi}=\mathbf{X}_+\mathbf{X}_-^{+}$. In the second case, the samples at $t_k$ form $\mathbf{C}\boldsymbol{\Phi}^k\mathbf{B}$, and right-multiplying by $\mathbf{B}^{+}$ gives $\mathbf{C}\boldsymbol{\Phi}^k$. Stacking these for $k=0,\ldots,n$ gives a matrix $\mathbf{O}$ of full column rank, and stacking them for $k=1,\ldots,n+1$ gives $\mathbf{O}\boldsymbol{\Phi}$, so $\boldsymbol{\Phi}=\mathbf{O}^{+}(\mathbf{O}\boldsymbol{\Phi})$. In both cases $\mathbf{A}=\log(\boldsymbol{\Phi})/\Delta t$ with the principal logarithm, and $(\vec\theta,\gamma)$ follows by applying $\mathbf{M}^{-1}$ to the $n\times n$ block of $\mathbf{A}$ and to $\vec\beta$. \hfill$\square$

As a side-note on the sample complexity,
once the identifiability conditions above are satisfied, one can increase the number of samples and gradually increase the precision of the identification method. There exist many theoretical works that pertain to the asymptotic \cite{JANSSON19981507, KNUDSEN200181} and finite sample properties \cite{oymak19, xin22} of such methods.
We stress that the identifiability statements above are noiseless: they
concern whether the parameters are determined by exact data. In practice the
system matrix is estimated from tomographic data, and estimating each
$x_j=\mathrm{Tr}[F_j\rho_{\rm S}]$ from $S$ copies incurs a standard error
$O(S^{-1/2})$, so that
$\lVert\vecop(\hat{\mathbf{A}})-\vecop(\mathbf{A})\rVert=O(n\,S^{-1/2})$ up
to the conditioning of the sampling map. Combined with
\eqref{eq:error_bound}, the error of the reconstructed parameters is therefore at most $\|\mathbf{M}^{-1}\| = 1/\sigma_{\min}(\mathbf{M})$ times this, i.e.\ $O\big(n\,S^{-1/2}/\sigma_{\min}(\mathbf{M})\big)$.\\


Condition (i) describes the two ways of fixing the frame of Sec.~\ref{sec:frame}. Under full tomography, $\mathbf{C}$ has full column rank, and $\mathbf{CM}^{(\text{lin})}$ has full rank either for a complete set of preparations or for a single initial state that is cyclic for $\mathbf{A}$ (Sec.~\ref{sec:examples}). With a complete set of preparations, $\mathbf{B}$ has full row rank, and a partial measurement suffices if $\mathbf{OM}^{(\text{lin})}$ has full rank. If neither $\mathbf{B}$ nor $\mathbf{C}$ has full rank but $\mathbf{CM}^{(\text{lin})}$ and $\mathbf{OM}^{(\text{lin})}$ do, the samples determine $(\boldsymbol{\Phi},\mathbf{B},\mathbf{C})$ only up to a similarity transform \cite{ho1966effective}, and $(\vec\theta,\gamma)$ is identifiable if the known $\mathbf{B}$ and $\mathbf{C}$ fix that transform. Without a bound on the spectrum, (ii) can be replaced by non-uniform sampling with irrational ratios $\Delta t_i/\Delta t_j$ \cite{Ding2009}. For a symmetric Kossakowski matrix, $\mathbf{T}_1$ and $\mathbf{T}_3$ \eqref{eq:T_3} take the place of $\mathbf{M}$ in the last step.

Since a closed quantum system can be viewed as an open quantum system with vanishing Lindblad terms in the GKSL equation, we may formulate the following  {statement} that applies to closed quantum systems. Result~1 applies to closed systems, for which $\gamma=0$ and $\vec\beta=0$, so no augmentation is needed and $\vecop\mathbf{A}=\mathbf{T}_1\vec\theta$. Unlike the open case, a single preparation never gives $\mathbf{CM}^{(\text{lin})}$ full rank for $N\ge 3$, so the unconstrained estimate of Result~1 needs several; if $\gamma=0$ is imposed in the estimation, fewer can suffice. For a closed system, $\mathbf{A}$ is the adjoint action of $H$ on $\mathfrak{su}(N)$, whose zero eigenvalue has geometric multiplicity at least $N-1$ (the traceless operators that are diagonal in the eigenbasis of $H$), so $\mathbf{A}$ has no cyclic vector and at least $N-1$ linearly independent preparations are needed, more if Bohr frequencies are degenerate. Furthermore, since $\mathbf{T}_{1}$ has full column rank, one may uniquely construct the Hamiltonian of the system.
\\

\subsubsection{Driven Open Quantum System Identification}
As derived in the previous section, controlled open quantum systems are represented by a BDS. Sontag \textit{et al.}~\cite{Sontag2009} showed that, for generic bilinear systems, the family $\mathcal{V}_{\alpha}$ of all pulses of fixed amplitude $\alpha \in \mathbb{R}$ and varying width $\tau \geq 0$ suffices for identification, whereas step inputs and single pulses do not. Mathematically, this class of input functions takes the form

\begin{equation}\label{eq:pulses_for_BDS}
    \begin{aligned}
        \vec{u}_{\tau,\alpha}(t) = \begin{cases}
            \alpha & 0 \leq t < \tau \\
            0 & t \geq \tau
        \end{cases}
        \end{aligned} {.}
\end{equation}

Note that $\tau = 0$ gives the zero input, i.e.\ the undriven experiment, while a constant input corresponds to $\tau\to\infty$. From this, $\mathcal{V}_{\alpha}$ can be constructed as
\begin{equation}\label{eq:non_constant_pulses}
    \begin{aligned}
    \mathcal{V}_{\alpha} = \{ \vec{u}_{\tau, \alpha} | \tau \geq 0 \},
    \end{aligned}
\end{equation}
with $\alpha \neq 0$.  {The main positive result of Ref.~\cite{Sontag2009} is that there exists a class of bilinear systems for which, given any pair $\sigma = (\mathbf A, \mathbf B, \mathbf C, \mathbf N)$ and $\widehat\sigma = (\widehat{\mathbf A}, \widehat{\mathbf B}, \widehat{\mathbf C}, \widehat{\mathbf N})$ in that class, I/O indistinguishability under $\mathcal{V}_{\alpha}$ implies I/O equivalence under all inputs,}
\begin{equation}
    \sigma \underrel{\mathcal{V}_{\alpha}}{\equiv} \widehat{\sigma} \iff \sigma \equiv \widehat{\sigma}.
\end{equation}
Hence $\mathcal{V}_{\alpha}$ is a sufficient input for identifying systems in this subset. Rather than verifying its hypotheses for the GKSL bilinear systems considered here, we prove identifiability directly for the pulse protocol we use (Result~2 below). Note that I/O equivalence is strongly related to the concept of similarity \cite{Sontag1998}, which was used for the identifiability conditions of closed quantum systems in \cite{Wang_2020}. For minimal bilinear systems, I/O equivalence holds if and only if the two realizations are related by a similarity transformation \cite{Isidori1973,sussmann1976existence}.

\noindent\textbf{Result 2 (driven systems).} Consider the augmented system \eqref{eq:standard_form} with known $\mathbf{N}_1,\ldots,\mathbf{N}_m$, a known initial state $\vec b$, and a known $\mathbf{C}$ of full column rank. Let at most one control be on at any time, with known amplitude $\alpha\neq0$, and let the inputs switch only at the sampling times $t_k=k\Delta t$, so that over each interval the generator is $\mathbf{A}+\alpha\mathbf{N}_c$ for a setting $c\in\{0,\ldots,m\}$, with $\mathbf{N}_0=0$. Suppose that every eigenvalue $\lambda$ of every $\mathbf{A}+\alpha\mathbf{N}_c$ satisfies $|\mathrm{Im}\,\lambda|<\pi/\Delta t$, for the true and all candidate parameters. Then $(\vec\theta,\gamma)\in\mathbb{R}^n\times\Herm(n)$ is identifiable from such inputs if and only if $\mathbf{CM}^{(\text{bi})}$ has full rank. In that case, $\mathbf{A}$ is determined by the samples of any finite set of pulse sequences for which the states sampled at the start of the intervals with some setting $c$ span $\mathbb{R}^{n+1}$, and such sets exist.

\noindent\emph{Proof.} Let $\boldsymbol{\Phi}_c=e^{(\mathbf{A}+\alpha\mathbf{N}_c)\Delta t}$, which maps a sample taken at the start of an interval with setting $c$ to the next sample. The states reachable at the sampling times are the vectors $\boldsymbol{\Phi}_{c_k}\cdots\boldsymbol{\Phi}_{c_1}\vec b$, and their span is the smallest subspace that contains $\vec b$ and is invariant under every $\boldsymbol{\Phi}_c$. Each $\boldsymbol{\Phi}_c$ is a power series in $\mathbf{A}+\alpha\mathbf{N}_c$, and conversely, by the eigenvalue condition, $(\mathbf{A}+\alpha\mathbf{N}_c)\Delta t$ is the principal logarithm of $\boldsymbol{\Phi}_c$ and hence a polynomial in it. A subspace is therefore invariant under all $\boldsymbol{\Phi}_c$ if and only if it is invariant under all $\mathbf{A}+\alpha\mathbf{N}_c$, i.e.\ under $\mathbf{A}$ and every $\mathbf{N}_k$, since $\mathbf{N}_0=0$ and $\alpha\neq0$. The span of the reachable states is thus the column space of $\mathbf{CM}^{(\text{bi})}$.
If $\mathbf{CM}^{(\text{bi})}$ has full rank, choose reachable states $\vec z_1,\ldots,\vec z_{n+1}$ that span $\mathbb{R}^{n+1}$ and, for each $j$, a pulse sequence that reaches $\vec z_j$ and then applies setting $c$ for one interval. The states follow from the samples via $\mathbf{C}^{+}$. Collecting the $\vec z_j$ and their successors as the columns of $\mathbf{X}_-^{(c)}$ and $\mathbf{X}_+^{(c)}$ gives $\mathbf{X}_+^{(c)}=\boldsymbol{\Phi}_c\mathbf{X}_-^{(c)}$ with $\mathbf{X}_-^{(c)}$ invertible, so $\boldsymbol{\Phi}_c$, then $\mathbf{A}=\log(\boldsymbol{\Phi}_c)/\Delta t-\alpha\mathbf{N}_c$, and then $(\vec\theta,\gamma)$ are determined as in Result~1.
Conversely, if $\mathbf{CM}^{(\text{bi})}$ does not have full rank, there is a nonzero matrix $\mathbf{E}$, which can be taken arbitrarily small, with zero last row that annihilates its column space. Then $\mathbf{A}+\mathbf{E}$ has the augmented form and coincides with $\mathbf{A}$ on that column space, which contains $\vec b$ and is invariant under $\mathbf{A}$ and all $\mathbf{N}_k$, so it produces the same samples for every input. For small $\mathbf{E}$ it also satisfies the eigenvalue condition, and since $\mathbf{M}$ is a bijection it corresponds to parameters different from $(\vec\theta,\gamma)$. \hfill$\square$

The inputs of Result~2 are concatenations of pulses from $\mathcal{V}_\alpha$ \eqref{eq:non_constant_pulses} with widths that are multiples of $\Delta t$, applied to one control at a time. Unlike in \cite{Sontag2009}, the $\mathbf{N}_k$ are known and the state is measured, so these inputs suffice without a genericity assumption, and each setting reduces to a linear problem as in Result~1. With full tomography, $\mathbf{OM}^{(\text{bi})}$ has full rank automatically. A single initial state suffices whenever $\mathbf{CM}^{(\text{bi})}$ has full rank, even if $\vec b$ is not cyclic for $\mathbf{A}$, as for $|00\rangle$ in Sec.~\ref{sec:examples}. When several settings have full-rank data, each determines $\mathbf{A}$; Sec.~\ref{sec:examples} averages the resulting estimates. For a symmetric Kossakowski matrix, $\mathbf{T}_1$ and $\mathbf{T}_3$ \eqref{eq:T_3} take the place of $\mathbf{M}$.

Result~2 applies to closed systems with $\gamma=0$ and $\vec\beta=0$, without augmentation and with $\mathbf{T}_1$ in place of $\mathbf{M}$.

Results~1 and~2 assume that $\mathbf{B}$ or $\mathbf{C}$ is known exactly (Sec.~\ref{sec:frame}). Under full tomography, as in the first case of Result~1 and in Result~2, preparation errors do not bias the estimate, since $\mathbf{A}$ is computed from measured states only; they enter only through the rank conditions. A miscalibrated measurement returns $\mathbf{T}\vec x$ instead of $\vec x$, where $\mathbf{T}=\id+\mathbf{E}$ has last row $(0,\ldots,0,1)$ because the constant coordinate is exact. The one-step maps are then estimated as $\mathbf{T}\boldsymbol{\Phi}_c\mathbf{T}^{-1}$, and the estimate of $\mathbf{A}$ from setting $c$ ($\mathbf{N}_c=0$ without control) satisfies
\begin{equation}\label{eq:spam_bias}
\hat{\mathbf{A}}-\mathbf{A}=[\mathbf{E},\mathbf{A}+\alpha\mathbf{N}_c]\,\mathbf{T}^{-1},
\end{equation}
so that, to first order in $\mathbf{E}$, the error in $(\vec\theta,\gamma)$ is at most $\|\mathbf{M}^{-1}\|$ times the norm of this commutator. The eigenvalues of each estimated generator $\log(\hat{\boldsymbol{\Phi}}_c)/\Delta t=\mathbf{T}(\mathbf{A}+\alpha\mathbf{N}_c)\mathbf{T}^{-1}$ are not affected at all; for the undriven setting they are the decay rates and Bohr frequencies, in line with the gauge-invariant quantities of gate-set tomography \cite{nielsen2021gate}. After subtraction of $\alpha\mathbf{N}_c$, the estimate from a driven setting is similar to $\mathbf{A}$ only if $\mathbf{E}$ commutes with $\mathbf{N}_c$.

\section{Example: a driven two-qubit system}\label{sec:examples}

We illustrate the conditions of Sec.~\ref{sec:sum_identification} on two coupled qubits with local dissipation, adapted from \cite{Zhang2015}. Since $\mathbf{M}$ is invertible, identifying the master equation is equivalent to identifying $\mathbf{A}$ and $\vec\beta$; the question is therefore which experiments produce data that determine them, and how accurately.
\begin{figure*}
\includegraphics[width=\textwidth]{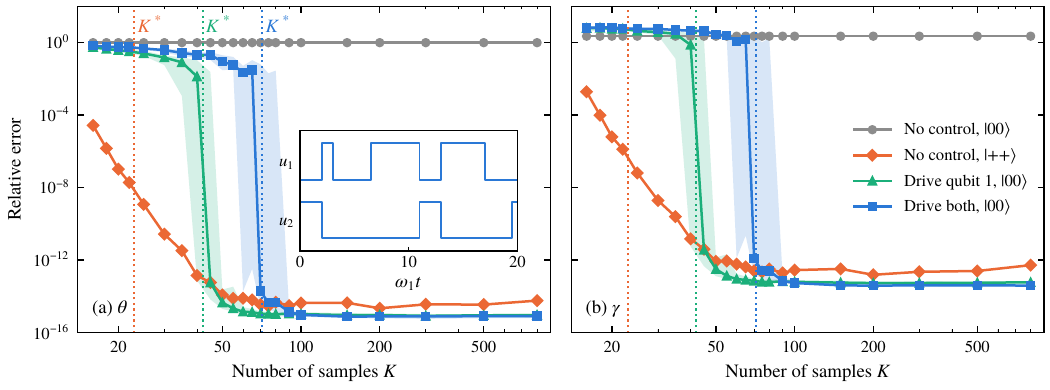}
\caption{Identifiability transition for the two-qubit model \eqref{eq:example_gksl}, from noise-free data. Relative error of (a) the Hamiltonian parameters, $\|\hat\theta-\theta\|/\|\theta\|$, and (b) the Kossakowski matrix, $\|\hat\gamma-\gamma\|_F/\|\gamma\|_F$, against the number of samples $K$, sampled every $\Delta t = 0.5/\omega_1$. Protocols: no control from $|00\rangle$ (circles) or $|{+}{+}\rangle$ (diamonds), and $\sigma_x$ pulses of amplitude $\omega_1$ from $|00\rangle$, with the setting, including off, drawn uniformly for segments of 1--5 samples, applied to qubit~1 only (triangles) or to one qubit at a time on either (squares). Markers show the median over 50 random pulse sequences, shaded bands the interquartile range, which is wide near $K^*$ because the $K$ at which full rank is reached varies between sequences. Dotted lines mark $K^*$, the median $K$ at which the data of every pulse setting reach rank $n+1 = 16$ ($K^* = 23$, 42 and 71); without control from $|00\rangle$ the rank saturates at 6. Inset: example pulse sequence when both qubits are driven. Parameters: $\omega_2 = 1.3\,\omega_1$, $\delta = 0.2\,\omega_1$, $(g_k^-,g_k^+,g_k^z) = (0.05,0.01,0.02)\,\omega_1$ for $k=1$ and $(0.03,0.02,0.04)\,\omega_1$ for $k=2$.}
\label{fig:controls}
\end{figure*}

\begin{figure*}
\includegraphics[width=\textwidth]{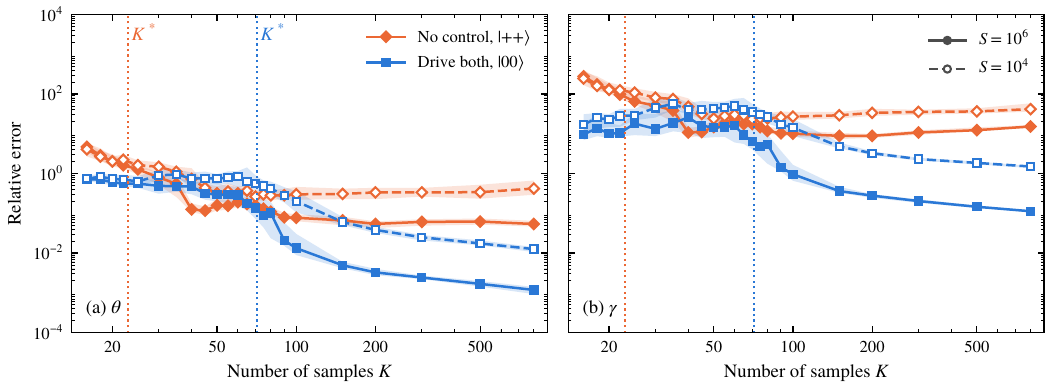}
\caption{Effect of shot noise, for the autonomous protocol from $|{+}{+}\rangle$ (diamonds) and the protocol with both qubits driven from $|00\rangle$ (squares); otherwise as in Fig.~\ref{fig:controls}. Each of the 15 Pauli expectation values at each sampling time is estimated from $S = 10^6$ (solid lines, filled markers) or $S = 10^4$ (dashed lines, open markers) projective measurements on fresh preparations. (a) Hamiltonian parameters $\theta$; (b) Kossakowski matrix $\gamma$, reconstructed over all 225 entries. Markers show the median over 50 pulse sequences, bands the interquartile range, and dotted lines $K^*$ from Fig.~\ref{fig:controls}.}
\label{fig:shots}
\end{figure*}
\subsection{Model}

The system evolves as
\begin{equation}\label{eq:example_gksl}
\begin{aligned}
\dot \rho= & -\im[H(t), \rho]+\sum_{k=1}^2 \frac{g_{k}^{z}}{2}\left(\sigma_z^k \rho \sigma_z^k-\rho\right) \\
& +\sum_{k=1}^2 2 g_k^{-}\left(\sigma_{-}^k \rho \sigma_{+}^k-\frac{1}{2} \sigma_{+}^k \sigma_{-}^k \rho-\frac{1}{2} \rho \sigma_{+}^k \sigma_{-}^k\right) \\
& +2 g_k^{+}\left(\sigma_{+}^k \rho \sigma_{-}^k-\frac{1}{2} \sigma_{-}^k \sigma_{+}^k \rho-\frac{1}{2} \rho \sigma_{-}^k \sigma_{+}^k\right),
\end{aligned}
\end{equation}
with $H(t)$ as in \eqref{control_hamiltonian} and drift Hamiltonian
\begin{equation}\label{eq:example_H}
    H_0 = \frac{\omega_1}{2}\sigma^{1}_{z} + \frac{\omega_2}{2}\sigma^{2}_{z} + \delta(\sigma^1_{+}\sigma^2_{-} + \sigma^1_{-}\sigma^{2}_{+}),
\end{equation}
where $\sigma_j^k$ acts as the Pauli operator $\sigma_j$ on qubit $k$ and as the identity on the other, and $\sigma_{\pm}^{k} = \frac{1}{2}(\sigma^k_x \pm \im\,\sigma^k_{y})$. We work in a frame rotating at the common frequency of the drives, so that $\omega_k$ are the detunings of the qubits from it, $\delta$ is an exchange coupling, and $g_k^{-}$, $g_k^{+}$, $g_k^{z}$ are the decay, excitation and dephasing rates of qubit $k$. The exchange term and the dissipators are unchanged by this transformation, since $\sigma_\pm^k$ only acquire phases that cancel in each term. The local form of the dissipators is appropriate because $\delta$ and the drive amplitude are small compared with the bare qubit frequencies \cite{hofer2017markovian}.

In the orthonormal basis $F_j = \sigma_a^1\sigma_b^2/2$ of $\mathfrak{su}(4)$, with $\sigma_0=\id$, the drift has four nonzero components: $\theta = \omega_1$ along $\sigma_z^1/2$, $\omega_2$ along $\sigma_z^2/2$, and $\delta$ along both $\sigma_x^1\sigma_x^2/2$ and $\sigma_y^1\sigma_y^2/2$. The Lindblad operators are not elements of this basis, so we expand them in it to obtain the Kossakowski matrix of \eqref{canonicalGKSL}. Because the dissipation is local, $\gamma$ is block diagonal, $\gamma = \gamma^{(1)}\oplus\gamma^{(2)}\oplus 0_{9\times9}$, with single-qubit blocks in the basis $(\sigma_x^k,\sigma_y^k,\sigma_z^k)/2$
\begin{equation}\label{eq:example_gamma}
\gamma^{(k)} = 2\begin{pmatrix}
 g_k^-+g_k^+ & \im\left(g_k^--g_k^+\right) & 0\\[3pt]
 -\im\left(g_k^--g_k^+\right) & g_k^-+g_k^+ & 0\\[3pt]
 0 & 0 & g_k^{z}
\end{pmatrix}.
\end{equation}
The ten nonzero entries encode the six rates. The imaginary off-diagonals are proportional to $g_k^- - g_k^+$ and vanish only in the unital case; they are also the source of $\vec\beta \neq 0$. Since $\mathbf{M}$ is invertible, Algorithm~\ref{alg:general_rec} recovers all $n + n^2 = 240$ real parameters from $(\mathbf{A},\vec\beta)$, without using the sparsity of $\theta$ or $\gamma$.

In the simulations we use units of $\omega_1$ and set $\omega_2 = 1.3$, $\delta = 0.2$, $(g_1^-,g_1^+,g_1^z) = (0.05, 0.01, 0.02)$ and $(g_2^-,g_2^+,g_2^z) = (0.03, 0.02, 0.04)$. The slowest relaxation rate of $\mathbf{A}$ is then $0.10$.

\subsection{Initial state, controls and sampling}

\emph{Initial state.} In the BDS \eqref{OQS_bilinear} the initial state is fixed, $\vec x(0) = \vec b$, so reachability depends on $\vec b$. Without control the trajectory $e^{\mathbf{A}t}\vec b$ stays in the span of $\{\mathbf{A}^k \vec b\}$, the column space of $\mathbf{CM}^{(\text{lin})}$ in \eqref{Linear_OM_CM} with $\mathbf{B} = \vec b$. For a computational-basis state such as $|00\rangle$ this space has dimension 6, not $n+1 = 16$. The reason is a symmetry: $H_0$ commutes with $\sigma_z^1 + \sigma_z^2$, and the jump operators only acquire a phase under rotations generated by it, so the undriven generator commutes with these rotations. A state that is invariant under them, such as $|00\rangle\langle 00|$, remains invariant at all times. It therefore stays in the span of $\id$, $\sigma_z^1$, $\sigma_z^2$ and $\sigma_z^1\sigma_z^2$ (the populations) and of $\sigma_x^1\sigma_x^2 + \sigma_y^1\sigma_y^2$ and $\sigma_x^1\sigma_y^2 - \sigma_y^1\sigma_x^2$ (the coherence between $|01\rangle$ and $|10\rangle$). No choice of sampling times enlarges this space. A state that breaks the symmetry, such as $|{+}{+}\rangle$ with $|{+}\rangle = (|0\rangle + |1\rangle)/\sqrt{2}$, gives $\rank\mathbf{CM}^{(\text{lin})} = 16$, so the autonomous system is identifiable from a single preparation, consistent with Result~1.

\emph{Controls.} We drive each qubit with $H_k = \sigma_x^k$ in \eqref{control_hamiltonian}, the form that a resonant transverse (microwave or radio-frequency) drive takes in the rotating frame within the rotating-wave approximation, with $\alpha$ setting the Rabi frequency. Since $\sigma_x^k$ does not commute with $\sigma_z^1 + \sigma_z^2$, it breaks the symmetry: starting from $|00\rangle$, the bilinear reachability condition \eqref{eq:BDS_reach_and_obs} holds with either drive alone, whereas a drive along $\sigma_z^k$ leaves the reachable space six-dimensional. One drive suffices because the coupling $\delta$ carries the excitation to the other qubit. The matrices $\mathbf{N}_k$ follow from the known $H_k$, so the drive adds no unknowns; its role is to make the data informative.

\emph{Pulses.} The inputs are drawn from $\mathcal{V}_\alpha$ \eqref{eq:non_constant_pulses} with $\alpha = \omega_1$: at any time at most one drive is on, and the setting, including off, is drawn uniformly at random for consecutive segments of 1--5 sampling intervals, so that off periods have the same distribution as pulses. The input therefore takes one of three values, $u = (0,0)$, $(\alpha,0)$ or $(0,\alpha)$, which we label as pulse settings $c = 0, 1, 2$, with $\mathbf{N}_{c=0} \equiv 0$. Pulses switch only at sampling times, so the generator is constant over each interval and the samples obey exactly
\begin{equation}\label{eq:example_step}
    \vec x_{k+1} = e^{(\mathbf{A} + \alpha\mathbf{N}_{c_k})\Delta t}\,\vec x_k ,
\end{equation}
where $c_k$ is the setting during the $k$-th interval and $\vec x$, $\mathbf{A}$, $\mathbf{N}_c$ are in the augmented form \eqref{eq:standard_form}.

\emph{Sampling and measurement.} The state is sampled at $K+1$ uniformly spaced times $t_k = k\Delta t$ with $\Delta t = 0.5$. Uniform sampling suffices: by \eqref{eq:alias_bound}, applied to the Hamiltonian of each setting, $2(\|H\|+\mathrm{Tr}\,\gamma)\,\Delta t\le 2.34<\pi$, so condition (ii) of Results~1 and~2 holds for all candidate parameters whose $\|H\|$ and $\mathrm{Tr}\,\gamma$ exceed the true values by less than a third. For the true parameters, $\max|\mathrm{Im}\,\lambda(\mathbf{A}+\alpha\mathbf{N}_c)|\,\Delta t=1.77$. Each sample is obtained by full tomography. Each of the 15 expectation values $x_j = \langle P_j\rangle/2$, with $P_j = 2F_j$ a Pauli string, is estimated from $S$ projective measurements, each on a fresh preparation evolved to $t_k$. The noise is then independent across samples and components, with variance $(1 - \langle P_j\rangle^2)/(4S)$.

\subsection{Estimation}

For each setting $c$ we collect the samples taken immediately before and after the intervals in which $c$ was applied, $\mathbf{X}_-^{(c)} = [\vec x_k]_{c_k = c}$ and $\mathbf{X}_+^{(c)} = [\vec x_{k+1}]_{c_k = c}$. By \eqref{eq:example_step}, $\mathbf{X}_+^{(c)} = e^{(\mathbf{A}+\alpha\mathbf{N}_c)\Delta t}\,\mathbf{X}_-^{(c)}$, and we estimate
\begin{equation}\label{eq:example_ls}
    \hat{\boldsymbol{\Phi}}_c = \mathbf{X}_+^{(c)}\big(\mathbf{X}_-^{(c)}\big)^{+}, \qquad
    \hat{\mathbf{A}} = \left\langle \frac{\log\hat{\boldsymbol{\Phi}}_c}{\Delta t} - \alpha\mathbf{N}_c \right\rangle_{c},
\end{equation}
where $^{+}$ is the pseudo-inverse and $\langle\cdot\rangle_c$ the average over settings. Below full rank we use the minimum-norm update $\hat{\boldsymbol{\Phi}}_c=\id+(\mathbf{X}_+^{(c)}-\mathbf{X}_-^{(c)})(\mathbf{X}_-^{(c)})^{+}$, which coincides with \eqref{eq:example_ls} at full rank and acts as the identity on directions the data have not reached. If $\hat{\boldsymbol{\Phi}}_c$ has eigenvalues on the negative real axis, as can happen with noisy data, its principal logarithm is not real \cite{higham2008functions}, and we use its real part. Algorithm~\ref{alg:general_rec} then returns $(\hat\theta,\hat\gamma)$. The least-squares problem has a unique solution if and only if $\rank\mathbf{X}_-^{(c)} = n+1 = 16$ for every $c$, that is, if the states visited under each setting span the augmented state space. This is the data condition of Result~2, required here for every setting rather than for one; it can only hold if $\mathbf{CM}^{(\text{bi})}$ has full rank, and it needs at least 16 samples per setting. We denote by $K^*$ the smallest $K$ at which it holds for all settings, as the median over random pulse sequences, treating singular values below $10^{-9}$ as zero. We report the relative errors $\|\hat\theta - \theta\|/\|\theta\|$ and $\|\hat\gamma - \gamma\|_F/\|\gamma\|_F$, as the median and interquartile range over 50 pulse sequences for each $K$.

\subsection{Results}

Figure~\ref{fig:controls} shows noise-free data for four protocols. Without control from $|00\rangle$ the error stays of order one for every $K$: the data span only six directions, so $\mathbf{A}$ is undetermined on the remaining ten. Without control from $|{+}{+}\rangle$ the data have full rank from $K = 16$ in exact arithmetic, but the error decreases gradually and reaches machine precision only at $K \approx 50$. A single relaxing trajectory gives nearly collinear samples, so the least-squares problem is badly conditioned; at the numerical tolerance used for $K^*$, full rank is reached at $K^* = 23$. Driven from $|00\rangle$, the error falls abruptly to machine precision at $K \approx K^*$, with $K^* = 42$ for one drive and $71$ for both. Our estimator needs 16 linearly independent samples in each setting, so its $K^*$ grows with the number of settings, from a lower bound of 32 and 48, respectively. By Result~2, full-rank data in a single setting already determine $\mathbf{A}$, so this growth reflects the per-setting estimator rather than the information in the data. The onset of exact reconstruction is thus predicted by the rank condition.

Figure~\ref{fig:shots} compares the autonomous protocol from $|{+}{+}\rangle$ with the driven protocol from $|00\rangle$ under shot noise. In the autonomous case the error stops decreasing after $K \approx 100$ and settles at $\approx 0.05$ for $\theta$ and $\approx 10$ for $\gamma$ at $S = 10^6$. The state relaxes to the steady state at rate $\approx 0.1$, so after a few relaxation times further samples add noise but no information about the dynamics. The drive instead keeps exciting the system, which is the persistent excitation of classical identification \cite{astrom1971system,ljung1971characterization}. Every new sample then remains informative, and the error keeps decreasing: at $K = 800$ it is $1.2\times10^{-3}$ ($S = 10^6$) and $1.3\times10^{-2}$ ($S = 10^4$) for $\theta$, and $0.11$ and $1.5$ for $\gamma$. A hundredfold increase in $S$ reduces the error by about an order of magnitude, consistent with the $O(S^{-1/2})$ scaling of Sec.~\ref{sec:autonomous_id}.  One-step maps without a real principal logarithm occur for the driven protocol only in the transition region, for up to two thirds of the maps at $S=10^4$ and for none at $K\ge 200$, and for the autonomous protocol at $S=10^6$ only for $K\le 80$. At $S=10^4$, however, they occur in most autonomous realizations at every $K$, so the corresponding curves in Fig.~\ref{fig:shots} indicate a failure of the estimator rather than an error level. Control is therefore not required for identifiability when a symmetry-breaking state can be prepared, but it makes accurate identification possible from a simple initial state.

The dissipative parameters are much harder to estimate than the Hamiltonian ones. The rates are one to two orders of magnitude smaller than $\omega_1$, and we reconstruct all 225 entries of $\gamma$ although only ten are nonzero. Restricting the reconstruction to the known support of $\gamma$, or imposing positivity of $\hat\gamma$, should reduce this error; we do not pursue this here.

\section{Algorithms and Numerical Stability}\label{sec:algorithms}

In this section, we provide pseudocode algorithms for
the master-equation parameter reconstruction of
Sec.~\ref{sec:parameter_reconstruction}, in the general case and in the
symmetric-Kossakowski case. The aim here is to further clarify the concepts developed for future practitioners to implement {post-identification} processing for reconstructing the GKSL master equation in terms of the original parameters. With these algorithms we also discuss numerical stability, focusing on the
conditioning of the matrices that must be inverted.

Firstly, we present Algorithm \ref{alg:general_rec} {for parameter reconstruction when} the form of the Kossakowski matrix has no discernible structure, which corresponds to the results in Sec.~\ref{sec:general_param_reconst}.
\\

\begin{algorithm}
    \caption{Parameter Reconstruction - General Case}\label{alg:general_rec}
    \KwIn{$\mathbf{A}, \vec{\beta}, \{F_{j}\}_{j=1}^{n}$}
    Construct the structure constants $f_{jkl}, g_{jkl}$ and $z_{jkl}$ as in \eqref{eq:antisymmtric_structure_const}, \eqref{eq:symmtric_structure_const}, \eqref{eq:LDS_matrices_and_vectors}\;
    Compute the dissipation tensor, $D_{lm}^{(j,k)}$, as in \eqref{eq:LDS_matrices_and_vectors}\;
    Construct the bijective map, $\Gamma$, as in \eqref{eq:bijectivemap}\;
    Compute $\mathbf{T}_{1}, \mathbf{T}_{2}$ as in \eqref{eq:reconstrct_matrices}\;
    Compute $\mathbf{M}$ as in \eqref{matrix_M}\;
    Compute \begin{equation*}\mathbf M^{-1}\begin{pmatrix}A_{\Gamma(1)}\\
\vdots\\
A_{\Gamma(\Ndim^{2})}\\
\beta_{1}\\
\vdots\\
\beta_{n}
\end{pmatrix}=\begin{pmatrix}\theta_{1}\\
\vdots\\
\theta_{\Ndim}\\
\gamma_{\Gamma(1)}\\
\vdots\\
\gamma_{\Gamma(\Ndim^{2})}
\end{pmatrix}
\end{equation*}
\KwOut{$\begin{pmatrix}\theta_{1},&
\cdots&
\theta_{\Ndim},&
\gamma_{\Gamma(1)},&
\cdots&
\gamma_{\Gamma(\Ndim^{2})}
\end{pmatrix}^{T}$}
\end{algorithm}

The numerical stability of this reconstruction is governed by $\mathbf{M}$ alone. Since $\mathbf{M}$ is built from structure constants, it is known exactly up to floating-point rounding, which is negligible compared with the estimation error of $(\mathbf{A},\vec\beta)$. The error in the parameter vector $\vec \xi = (\vec\theta,\vec\gamma)$ therefore obeys
\begin{equation}\label{eq:error_bound}
\|\vec \xi-\hat{\vec \xi}\| \le \|\mathbf{M}^{-1}\|\,
\big\|\big(\vecop(\hat{\mathbf{A}})-\vecop(\mathbf{A}),\ \hat{\vec\beta}-\vec\beta\big)\big\|,
\end{equation}
where $\|\mathbf{M}^{-1}\| = 1/\sigma_{\min}(\mathbf{M})$, and the relative error is at most $\kappa(\mathbf{M})$ times that of $(\mathbf{A},\vec\beta)$. For $\mathfrak{su}(2)$, $\mathfrak{su}(3)$ and $\mathfrak{su}(4)$ we find $\sigma_{\min}(\mathbf{M}) = 1$, $0.631$ and $0.526$, and $\sigma_{\max}(\mathbf{M}) = N$ in each case ($\kappa = 2.0$, $4.8$ and $7.6$), independent of the orthonormal basis, so for two qubits the reconstruction amplifies absolute errors by at most a factor of 1.9.

A similar analysis can be produced for the case of Algorithm \ref{alg:symm_rec}, replacing $\mathbf{M}$ with $\mathbf{T}_{1}$ or  $\mathbf{T}_{3}$ and $\mathbf{A}$ for $\mathbf{A}^{(l)}$ or $\mathbf{A}^{(d)}$ respectively.

Measurement noise does not affect the rank of $\mathbf{M}$ as this is built from structure constants only and is exact, so noise enters solely through $(\mathrm{vec}\,\hat{\mathbf{A}},\hat{\vec\beta})$ and is amplified by at most $\|\mathbf{M}^{-1}\|=1/\sigma_{\min}(\mathbf{M})$. Noise does, however, enter the estimation of $\mathbf{A}$, where the recorded states act as regressors: the estimation error scales as $S^{-1/2}/\sigma_{\min}(\mathbf{X}_-)$, with $\mathbf{X}_-$ the matrix of states preceding each step (Sec.~\ref{sec:examples}), which reduces to $\sigma_{\min}(\mathbf{B})$ for independently prepared states. The rank conditions are therefore meaningful only if $\sigma_{\min}(\mathbf{X}_-)$ is well above the shot-noise level. In Sec.~\ref{sec:examples}, a single autonomous trajectory from $|{+}{+}\rangle$ has full rank but a $\sigma_{\min}$ more than an order of magnitude smaller than the driven data, and its error stops decreasing with $K$. This is the quantitative form of persistent excitation~\cite{astrom1971system}.

Next, we produce Algorithm \ref{alg:symm_rec} covering the results of Sec.~\ref{sec:param_reconst_sym}, in which case the Kossakowski matrix is known to be symmetric.

\begin{algorithm}[h]
\caption{Parameter Reconstruction - Symmetric Kossakowski Matrix}\label{alg:symm_rec}
\KwIn{$\mathbf{A}, \{F_{j}\}_{j=1}^{n}$}
Construct the structure constants $f_{jkl}, g_{jkl}$ and $z_{jkl}$ as in \eqref{eq:antisymmtric_structure_const}, \eqref{eq:symmtric_structure_const} and \eqref{eq:LDS_matrices_and_vectors}\;
Compute the symmetric dissipation tensor, $\tilde{D}_{lm}^{(j,k)}$, using \eqref{symmeytric_a_d}\;
Construct $\Gamma_{\rm sym}$ and $\mathbf{P}_{\rm sym}$ as defined above \eqref{eq:T_3}\;
Obtain $\mathbf{A}^{(l)}, \mathbf{A}^{(d)}$ from the symmetric decomposition of $\mathbf{A}$ \eqref{eq:toeplitz_Decomposition}\;
Compute $\mathbf{T}_{1}$, $\mathbf{T}_{3}$ as in \eqref{eq:reconstrct_matrices} and \eqref{eq:T_3}\;
Calculate the pseudoinverses $\mathbf{T}_{1}^{+}$ and $\mathbf{T}_{3}^{+}$\;
Compute $\displaystyle \mathbf{T}_{3}^{+}\begin{pmatrix}A_{\Gamma(1)}^{(d)}\\ \vdots\\ A_{\Gamma(n^{2})}^{(d)}\end{pmatrix}=\begin{pmatrix}\gamma_{\Gamma_{\rm sym}(1)}\\ \vdots\\  \gamma_{\Gamma_{\rm sym}(n(n+1)/2)}\end{pmatrix}$\;
Compute $\displaystyle \mathbf{T}_{1}^{+}\begin{pmatrix}A_{\Gamma(1)}^{(l)}\\ \vdots\\ A_{\Gamma(n^{2})}^{(l)}\end{pmatrix}=\begin{pmatrix}\theta_{1}\\ \vdots\\ \theta_{n}\end{pmatrix}$\;
  \textbf{Output:} $\left(\theta_1,\ldots,\theta_n,\ \gamma_{\Gamma_{\rm sym}(1)},\ldots,\gamma_{\Gamma_{\rm sym}(n(n+1)/2)}\right)^{T}$
\end{algorithm}

\FloatBarrier
\section{Conclusion}\label{sec:conclusions}
In this work, we presented a unified view of open quantum system identifiability by connecting GKSL master equations to linear and bilinear dynamical system formulations. Utilizing the concepts of controllability, observability, and minimality, we have shown how discrete-time measurement data can reveal both Hamiltonian and decoherence parameters. Specifically, time-independent controlled quantum systems align with bilinear models, whereas autonomous systems correspond to linear ones. By the uniqueness part of the GKS theorem, made explicit here through the structure-constant matrix $\mathbf{M}$, the master-equation parameters are always uniquely determined by the system matrices, so that identifiability reduces to that of the corresponding linear or bilinear system. The quantum-specific constraints, positivity of $\gamma$ and known structure, are then what distinguishes the problem from classical identification.

The two-qubit example shows the reconstruction at work on simulated tomographic data. Exact reconstruction sets in once the data matrix of every pulse setting reaches full rank, so the rank conditions predict how many samples are needed. Without control, the outcome depends on the initial state: a symmetric state such as $|00\rangle$ excites only six of the sixteen directions of the augmented state space, whereas a symmetry-breaking state such as $|{+}{+}\rangle$ suffices in principle. Under shot noise, however, the autonomous estimate stops improving once the system has relaxed, while driving keeps the data informative and the error decreasing with the number of samples. Because $\mathbf{M}$ is invertible, with the rank deficiency of $\mathbf{T}_{2}$ compensated by $\vec\beta$ (Sec.~\ref{sec:general_param_reconst}), the full Kossakowski matrix is recovered without restriction to its support; the dissipative parameters nonetheless remain the hardest to estimate, which motivates exploiting their sparsity and positivity. We also discussed cases in which only partial tomography or constrained controls are available, noting that sampling protocols must adapt accordingly as system dimensionality grows.

Our work bridges quantum and classical methods, leveraging multi-rate sampling, subspace identification and classical estimation techniques. While we touched upon embedding and non-Markovian extensions, a deeper unification of noise embedding in system identification is an important future direction.

{We close with a remark on scale. The coherence-vector representation is
exact but exponentially large: for a register of $q$ qubits, $n=4^{q}-1$, so
that $\mathbf{T}_{3}$ has dimensions $n^{2}\times n(n+1)/2$, that is
$3969\times2016$ already at $q=3$ and $65025\times32640$ at $q=4$. The
framework as presented is therefore a tool for the careful characterization of
few-qubit devices---precisely the regime in which gate calibration and noise
certification are performed---rather than a method for many-body systems. Two
ingredients developed above indicate the route beyond it: the accessible-set
reduction \eqref{eq:accessible_set}, which restricts the dynamics to the
subspace actually probed by the available observables, and restriction of the reconstruction to the sparse support of $\gamma$, which replaces the full parameter space by the sparse set that locality and symmetry
permit. Combining the two with locality-aware tomography is, in our view, the
natural next step.}

Overall, these results underscore the power of classical system-identification theory for analyzing and controlling open quantum systems: it fully applies whenever the dynamics can be approximated by GKSL equations. As the field advances toward complex, multi-qubit architectures and real-time feedback control, {we expect frameworks of this kind (and the identifiability questions they make precise) to become an increasingly standard part of device characterization}.

\bigskip
\section*{Data Availability}

The code reproducing Figs.~\ref{fig:controls} and \ref{fig:shots} is available in \cite{parvaiz2026numerics}.

\section*{Acknowledgments}
We would like to thank Denys I. Bondar, Zakhar Popovych, and Kurt Jacobs for extensive discussions of open quantum systems and their identification, while working on \cite{popovych2024quantumopenidentificationglobal}.
J.M. would also like to thank Jiri Vala.
The work of W.P., A.W. and J.M. has been supported by the
Czech Science Foundation (23-07947S). The work of G.K. has been partially supported by project MIS 5154714 of the National Recovery and Resilience Plan Greece 2.0 funded by the European Union under the NextGenerationEU Program.

\section*{Author Contributions}

J.M. and G.K. proposed the project and supervised this work. The background material on open quantum system and system identification was researched by W.P., A.W. and J.A. W.P. formulated the controlled open quantum system Lindbladian as a bilinear dynamical equation.  W.P., A.W. and J.A. developed the theory on the conditions of parameter reconstruction. The code for the two-qubit example was developed by W.P.

\section*{Competing interests}
This paper was prepared for information purposes
and is not a product of HSBC Bank Plc. or its affiliates.
Neither HSBC Bank Plc. nor any of its affiliates make
any explicit or implied representation or warranty, and
none of them accept any liability in connection with
this paper, including, but not limited to, the completeness,
accuracy, reliability of information contained herein and
the potential legal, compliance, tax, or accounting effects
thereof.

\appendix

\section{Derivation of the bilinear form}\label{ap:derivation_LDS}

Let us derive the bilinear form
for the controlled GKSL dynamics, as given in equation \eqref{eq:BLDS_lindbladian}. The steps in this proof follow closely those presented in Alicki and Lendi \cite{Alicki_Lendi_2014, Chruscinski2017}, which show how an uncontrolled system can be written as a linear ordinary differential equation. We only differ by introducing the control which leads to the bilinear term.

The starting point for the derivation of the bilinear dynamical system from the controlled open quantum systems, is to write the update equation for the elements of the state vector
\begin{equation}
    \begin{aligned}
        \dot{x}_{i} =& \mathrm{Tr}\left[F_{i} \dot\rho_{\rm{S}}(t)\right] \\
        =& \underbrace{-\im\mathrm{Tr}\left[F_{i} \left[H_{0}, \rho_{\rm{S}}(t)  \right]\right]}_{\dot{x}_{i}^{(l)}}
        \\&+ \underbrace{\frac{1}{2}\sum_{kl}\gamma_{kl}\mathrm{Tr}\left[F_{i}\left(\left[F_{k}, \rho_{\rm{S}}(t)F_{l}\right] + [F_{k}\rho_{\rm{S}}(t), F_{l}]\right) \right]}_{\dot{x}_{i}^{(d)}}
        \\&\underbrace{-\im\sum_{k=1}^{m}u_{k}(t)\mathrm{Tr}\left[F_{i}\left[H_{k}, \rho_{\rm{S}}(t) \right] \right]}_{\dot{x}_{i}^{(c)}}.
    \end{aligned}
\end{equation}

In the above, we have used the linear mapping properties of the trace and commutator to separate the contribution to the evolution by the Liouville-von Neumann, dissipation and control terms and distinguish these using underbraces. We will tackle each of the terms below and also commonly make use of a trace identity
\begin{equation}\label{eq:trace_identity}
    \mathrm{Tr}\left[A\left[B,C\right] \right] = \mathrm{Tr}\left[\left[A, B\right]C\right],
\end{equation}
which is true due to the cyclic property of the trace operator ($\mathrm{Tr}[ABC] = \mathrm{Tr}\left[CAB\right]$).

\subsection{Liouville-von Neumann Term}

We begin with the contribution from the closed system dynamics and substitute the decomposition of the system Hamiltonian \eqref{eq:decomposition}.

\begin{equation}
    \begin{aligned}
        \dot{x}^{(l)}_{i} &= -\im\sum_{k}\theta_{k}\mathrm{Tr}\left[ F_{i}\left[F_{k}, \rho_{\rm{S}}(t)\right] \right] \\
        &= -\im\sum_{k}\theta_{k}\mathrm{Tr}\left[ \left[F_{i},F_{k}\right]\rho_{\rm{S}}(t) \right]
        \\
        &= \sum_{j,k}\theta_{k}f_{kji}\mathrm{Tr}\left[F_{j}\rho_{\rm{S}}(t)\right] \\
        &= \sum_{j,k}\theta_{k}f_{kji}x_{j}.
    \end{aligned}
\end{equation}

We have used the identity \eqref{eq:trace_identity} to arrive at the second line and then further the identity of the antisymmetric structure constants immediately after (equation \eqref{eq:antisymmtric_structure_const}). Hence, recognizing that the final expression has matrix-vector form
\begin{equation}
    \dot{\vec{x}}^{(l)} = \mathbf{A}^{(l)}\vec{x}.
\end{equation}

\subsection{Dissipation Terms}

Next, we derive the contribution of the environment induced dissipation, $\dot{x}^{(d)}_{i}$. Firstly, by taking the addition of equations \eqref{eq:antisymmtric_structure_const} and \eqref{eq:symmtric_structure_const}, we obtain the relation
\begin{align}
    F_{i}F_{j} = \frac{1}{N}I\delta_{ij} +\frac{\im}{2}\sum_{k}\bar{z}_{ijk}F_{k},
\end{align}
with $z_{ijk} = f_{ijk} + \im g_{ijk}$ as a combined structure constant. Hence
\begin{equation*}
    \begin{aligned}
\frac{1}{2}F_{l}[F_i,F_{k}]  &=  \frac{\im}{2}F_{l}\sum_{m}f_{ikm}F_{m} \\
&= \frac{\im I}{2N}f_{ikl} - \frac{1}{4}\sum_{mj}f_{ikm}\bar{z}_{lmj}F_{j},
    \end{aligned}
\end{equation*}
and similarly
\begin{equation*}
    \begin{aligned}
\frac{1}{2}[F_l,F_{i}]F_{k}
    &= \frac{\im I}{2N}f_{ikl}  - \frac{1}{4}\sum_{mj}f_{lmi}z_{kmj}F_{j}.
    \end{aligned}
\end{equation*}

Using these terms we get

\begin{equation}
    \begin{aligned}
        \dot{x}_{i}^{(d)} &= \sum_{kl}\gamma_{kl}\mathrm{Tr}[(\frac{\im}{2N}f_{ikl} -\frac{1}{4}\sum_{mj}f_{ikm}\bar{z}_{lmj}F_{j}  +\frac{\im}{2N}f_{ikl} \\ & - \frac{1}{4}\sum_{mj}f_{lmi}z_{kmj}F_{j})\rho_{\rm{S}}] \\
    &= -\frac{1}{4}\sum_{klmj}\gamma_{kl}(f_{ikm}\bar{z}_{lmj} + f_{lmi}z_{kmj})\mathrm{Tr}[F_{j}\rho_{\rm{S}}]  \\ & + \frac{\im}{N}\sum_{kl}\gamma_{kl}f_{ikl}\mathrm{Tr}[\rho_{\rm{S}}]\\
    &= - \frac{1}{4}\sum_{klmj}\gamma_{kl}(f_{ikm}\bar{z}_{lmj} + f_{lmi}z_{kmj})x_{j} +\frac{\im}{N}\sum_{k,l}\gamma_{kl}f_{ikl}\\
    =& -{\sum_{j}}\sum_{kl}\gamma_{kl}D^{(i,j)}_{k,l}x_{j} +\frac{\im}{N}\sum_{k,l}\gamma_{kl}f_{ikl}
    \end{aligned}
\end{equation}
which are the elements of $\mathbf{A}^{(d)}_{ij}$ and $\beta_{i}$ respectively, and $D^{(i,j)}_{k,l}$ was introduced to condense notation.

\subsection{Control Term}\label{ap:control_term}

Finally, we consider the dynamics due to the control term $-\im\sum_{k=1}^{m}u_k(t)[H_k,\rho_{\rm S}]$ in \eqref{canonicalGKSL}, with $H_k=\sum_{r}h_{kr}F_r$ as in \eqref{reformulated_control_hamiltonian}. The derivation is the same as for the Hamiltonian part, again using \eqref{eq:trace_identity}:
\begin{equation}
    \begin{aligned}
       \dot x^{(c)}_{i}(t) &= -\im\sum_{k=1}^{m}u_{k}(t)\, \mathrm{Tr}\left(F_{i}\left[H_{k}, \rho_{\rm{S}}\right]\right) \\
        &= -\im\sum_{k=1}^{m}u_{k}(t)\sum_{r}h_{kr}\, \mathrm{Tr}\left( \left[ F_{i}, F_{r}\right]\rho_{\rm{S}} \right) \\
        &= \sum_{k=1}^{m}u_{k}(t)\sum_{r,l}h_{kr}f_{irl}\,\mathrm{Tr}\left(F_{l}\rho_{\rm{S}}\right)\\
        &= \sum_{k=1}^{m}u_{k}(t)\sum_{l}\Big(\sum_{r}h_{kr}f_{irl}\Big)x_{l}.
    \end{aligned}
\end{equation}
Hence
\begin{equation}
    \begin{aligned}
    \dot{\vec{x}}^{(c)}(t) &= \sum_{k=1}^{m}\mathbf{N}_{k}u_{k}(t)\vec{x}(t),\\
    (\mathbf{N}_k)_{il} &= \sum_{r}h_{kr}f_{irl} = -\sum_{r}h_{kr}f_{ilr},
    \end{aligned}
\end{equation}
in agreement with \eqref{eq:LDS_matrices_and_vectors}. The bilinear matrices are therefore fixed by the structure constants and the coefficients of the known control Hamiltonians. Since $[H_k,\id]=0$, the control term has no affine part, which is why $\mathbf{N}_k$ is padded with zeros in \eqref{eq:standard_form}. This completes the derivation.

\section{Proof of General Reconstruction result}\label{ap:proof_thm_reconstruct_general}

    After a successful identification procedure, we obtain the system matrix $\mathbf{A}$, and the vector $\vec{\beta}$. Vectorizing the whole problem by considering  $\phi$ and $\hat{\phi}$ leads to
   \begin{equation}\label{eq:ap_T_general_eq}
    \begin{aligned}
        \phi\left(\vec{\theta},\vec{\gamma}\right)=\mathbf{A}^{(l)}(\vec{\theta})+\mathbf{A}^{(d)}(\vec{\gamma}) \text{ and }\hat{\phi}\left(\vec{\gamma}\right)=\vec{\beta}\Longleftrightarrow
    \end{aligned}
\end{equation}

\begin{equation}\begin{pmatrix}A_{\Gamma(1)}\\
\vdots\\
A_{\Gamma(\Ndim^{2})}\\
\beta_{1}\\
\vdots\\
\beta_{n}
\end{pmatrix}=\mathbf M\begin{pmatrix}\theta_{1}\\
\vdots\\
\theta_{\Ndim}\\
\gamma_{\Gamma(1)}\\
\vdots\\
\gamma_{\Gamma(\Ndim^{2})}
\end{pmatrix},
\end{equation}
where
\begin{equation}
\mathbf{M}=\left[\begin{array}{cc}
\mathbf{T}_{1} & \mathbf{T}_{2}\\
0 & -\frac{\im}{N} \mathbf{T}_1^{\top}
\end{array}\right]\in\mathbb{C}^{\left(n+n^{2}\right)\times\left(n+n^{2}\right)}.
\end{equation}
Here, $\mathbf{T}_1$ and $ \mathbf{T}_2$ are the matrices defined in \eqref{eq:reconstrct_matrices}.
Hence $\vec\theta$ and $\gamma$ are uniquely determined by $\mathbf{A}$ and $\vec\beta$ whenever $\mathbf{M}$ is invertible, which holds for every orthonormal basis of traceless Hermitian matrices. Suppose $\mathbf{M}(\vec\theta,\gamma)^{\top}=0$ for complex $\vec\theta$ and $\gamma$, i.e.\ $\mathbf{A}=0$ and $\vec\beta=0$. Then the generator $\mathcal{L}$ in \eqref{canonicalGKSL} annihilates $\id$ and every $F_k$, and hence vanishes identically. Writing
$\mathcal{L}(\rho)=G\rho+\rho G'+\sum_{j,k=1}^{n}\gamma_{jk}F_j\rho F_k$ with $G,G'=\mp \im H-\tfrac12\sum_{j,k}\gamma_{jk}F_kF_j$,
and expanding $G$ and $G'$ in $\{F_0,\dots,F_n\}$ using $\id=\sqrt{N}F_0$, $\mathcal{L}$ takes the form $\sum_{a,b=0}^{n}c_{ab}F_a\rho F_b$ with $c_{jk}=\gamma_{jk}$ for $j,k\geq1$ and $c_{a0}=\sqrt{N}\,\mathrm{Tr}[F_aG]$ for $a\geq1$. Since $\mathrm{vec}(F_a\rho F_b)=(F_b^{\top}\otimes F_a)\,\mathrm{vec}(\rho)$ and $\{F_b^{\top}\otimes F_a\}_{a,b=0}^{n}$ is a basis of $\mathbb{C}^{N^2\times N^2}$, $\mathcal{L}\equiv0$ forces $c_{ab}=0$ for all $a,b$. Hence $\gamma=0$, so that $G=-\im H$ and $c_{a0}=-\im\sqrt{N}\,\theta_a=0$, i.e.\ $\vec\theta=0$. Thus $\ker\mathbf{M}=\{0\}$; this is the uniqueness part of the GKS theorem~\cite{gorini1976completely,Hall2014}. Injectivity over $\mathbb{C}$ implies injectivity on the real subspace $\mathbb{R}^n\times\Herm(n)$ of physical parameters. Since this subspace and the target $\mathbb{R}^{n^2}\times\mathbb{R}^n$ have the same real dimension $n+n^2$, $\mathbf{M}$ restricted to it is a real-linear bijection.

\section{Proof of the symmetric-Kossakowski result}\label{ap:proof_thm_reconstruct_diagonal}
Since the symmetric/antisymmetric decomposition is unique and
\begin{equation}
\mathbf{A}=\mathbf{A}^{(l)}+\mathbf{A}^{(d)},
\end{equation}
where $\mathbf{A}^{(l)}$ is always antisymmetric and $\mathbf{A}^{(d)}$ is symmetric due to the Kossakowski matrix being symmetric, we may directly compute $\mathbf{A}^{(d)}$ and $\mathbf{A}^{(l)}$ from $\mathbf{A}$ by
\begin{equation}
\mathbf{A}^{(d)}=\frac12(\mathbf A+\mathbf{A}^{\top}),\quad \mathbf{A}^{(l)}=\frac12(\mathbf A-\mathbf{A}^\top).
\end{equation}

Next, we write down the system of equations \eqref{eq:LDS_matrices_and_vectors}, using a vectorization that utilizes $\Gamma$, $\mathbf{T}_1$ and $\mathbf{T}_2$ as
\begin{equation}
\begin{aligned}
\begin{pmatrix}A_{\Gamma(1)}^{(l)}\\
\vdots\\
A_{\Gamma(n^{2})}^{(l)}
\end{pmatrix}=&\mathbf{T}_{1}\begin{pmatrix}\theta_{1}\\
\vdots\\
\theta_{n}
\end{pmatrix} \\
\begin{pmatrix}A_{\Gamma(1)}^{(d)}\\
\vdots\\
A_{\Gamma(n^{2})}^{(d)}
\end{pmatrix}=&\mathbf{T}_{2}\begin{pmatrix}\gamma_{\Gamma(1)}\\
\vdots\\
\gamma_{\Gamma(n^{2})}
\end{pmatrix}
\end{aligned}
\end{equation}
Using the symmetry of the Kossakowski matrix once more, we define a basis of the space in which $\gamma_{ij}=\gamma_{ji}$ if $i \neq j$, which results in the second of the two equations taking the form
\begin{equation}
\begin{pmatrix}A_{\Gamma(1)}^{(d)}\\
\vdots\\
A_{\Gamma(n^{2})}^{(d)}
\end{pmatrix}=\mathbf{T}_{3}\begin{pmatrix}\gamma_{\Gamma(1)}\\
\vdots\\
\gamma_{\Gamma(\frac{n\left(n+1\right)}{2})}
\end{pmatrix},
\end{equation}
where $\Gamma$ denotes the transformation, which assigns indices.

From the first equality, we can see that a sufficient condition for recovery of $\vec{\theta}$ and $\vec{\gamma}$
is
\begin{equation}
\begin{pmatrix}A_{\Gamma(1)}^{\left(d\right)}\\
\vdots\\
A_{\Gamma(n^{2})}^{\left(d\right)}
\end{pmatrix}\in\ran\mathbf{T}_{3},\begin{pmatrix}A_{\Gamma(1)}^{\left(l\right)}\\
\vdots\\
A_{\Gamma(n^{2})}^{\left(l\right)}
\end{pmatrix}\in\ran\mathbf{T}_{1}
\end{equation}

Moreover, $\mathbf{T}_{1}$ and $\mathbf{T}_{3}$ both have full column rank (Sec.~\ref{sec:param_reconst_sym}), so the recovery is unique.


\section{Structure constants of $\mathfrak{su}(2^{q})$}\label{ap:struct_const_orthog}

In the generalized Pauli basis, $F_a = P_a/\sqrt{N}$, where each $P_a$ is a tensor product of Pauli matrices and identities, not all identities, and $N = 2^q$. Any two Pauli strings either commute or anticommute, and their product is again a Pauli string up to a phase, $P_aP_b = \zeta_{ab}P_{c}$ with $c = c(a,b)$ and $\zeta_{ab}\in\{\pm1,\pm\im\}$. Hence:
\begin{itemize}
\item if $P_a$ and $P_b$ commute, then $[F_a,F_b]=0$ and $\{F_a,F_b\} = 2\zeta_{ab}P_{c}/N$, so $f_{abl}=0$ for all $l$, and $g_{abl}\neq 0$ only for $l=c$ (for no $l$ if $a=b$, since then $P_aP_b=\id$);
\item if they anticommute, then $\{F_a,F_b\}=0$ and $[F_a,F_b] = 2\zeta_{ab}P_c/N$, so $g_{abl}=0$ for all $l$, and $f_{abl}\neq0$ only for $l=c$.
\end{itemize}
Therefore, for every pair $(a,b)$ there is at most one index $l$ with $f_{abl}\neq0$ or $g_{abl}\neq0$, and $f_{ab\,\cdot}$ and $g_{ab\,\cdot}$ are never both nonzero. The property depends on the basis: it fails, for example, for the Gell-Mann basis of $\mathfrak{su}(3)$, where $f_{453}=1/2$ and $f_{458}=\sqrt{3}/2$ in the standard normalization.

\renewcommand{\bibsection}{\section*{ References}}
\bibliography{apssamp}

\end{document}